\documentclass[prd,twocolumn,superscriptaddress,altaffilletter,showpacs,nofootinbib]{revtex4}
\usepackage{amssymb,amsmath}
\usepackage{comment}
\usepackage{graphicx}
\usepackage{csquotes}
\usepackage{tikz}
\usepackage{xcolor}
\usetikzlibrary{snakes}
\usepackage{lipsum, babel}
\DeclareUnicodeCharacter{2212}{\textendash}

\newcommand{\proofof}[1]{\noindent {\bf Proof of #1. }}
\newtheorem{lemma}{Lemma}
\newtheorem{proposition}{Proposition}

\begin{document}

\title{Accretion Disks in Schwarzschild-MOG and Kerr-MOG Backgrounds: \\
MOG Parameter in terms of Observational Quantities}
\author{José Miguel Rojas}
\email{jrojas84@uasd.edu.do}
\affiliation{Mesoamerican Centre for Theoretical Physics, Universidad Autónoma de Chiapas,
Carretera Zapata Km. 4, Real del Bosque, 29040, Tuxtla Gutiérrez, Chiapas, México}
\affiliation{Instituto de Física, Universidad Autónoma de Santo Domingo, Av. Alma Mater, Santo Domingo 10105, Dominican Republic}

\author{Mehrab Momennia}
\email{mehrab.momennia@umich.mx}
\affiliation{Facultad de Ciencias F\'isico-Matem\'aticas, Universidad Michoacana de San Nicol\'as de Hidalgo,
Ciudad Universitaria, 58040, Morelia, Michoac\'an, M\'exico}

\date{\today }

\begin{abstract}
We apply a general relativistic framework to static and rotating black hole solutions in Scalar–Tensor–Vector Gravity or modified gravity (MOG). Our results yield exact analytic, closed-form relations expressing the mass $M$, the MOG coupling parameter $\alpha$, and the distance $D$ of a Schwarzschild–MOG black hole in terms of a minimal set of directly measurable elements of the accretion disk: the total frequency shift, the telescope aperture angle, and the \emph{redshift rapidity}. The resulting expressions are derived for particles close to the midline and line of sight, where the \emph{redshift rapidity} is treated as a relativistic invariant encoding the evolution of the frequency shift with respect to the emitter’s proper time in MOG spacetime. We further extend the formalism to the rotating Kerr–MOG geometry and obtain corresponding relations that determine the rotation parameter $a$ jointly with $M$, $\alpha$, and $D$ on the midline. In the rotating background, we introduced the \emph{redshift acceleration} (general-relativistic version of jerk) to disentangle the spacetime parameters. Crucially, the explicit appearance of $\alpha$ in these formulas enables direct empirical estimation of this parameter, thereby providing a means to test for departures from standard general relativity. The present study focuses on the geodesic motion of massive test particles revolving around Schwarzschild/Kerr–MOG  black holes, and the previous results obtained in the standard Schwarzschild/Kerr backgrounds are recovered in the limit $\alpha \to 0$.
The derived expressions are concise and suitable for incorporation into black hole parameter-estimation pipelines.

\vskip3mm

\end{abstract}

\maketitle

\section{Introduction}

Black holes began as bold mathematical solutions to Einstein’s field equations and they are now firmly established inhabitants of the cosmos. Foundational results in general relativity demonstrated that under broad physical conditions, gravitational collapse inevitably produces spacetime singularities~\cite{Penrose1965}. In parallel, direct detections of gravitational waves from black hole binary mergers~\cite{Abbott2016} and horizon-scale imaging of M87*~\cite{EHT2019I} have transformed black holes from theoretical constructs into empirical facts. Within general relativity theory, stationary black holes are remarkably economical: uniqueness (no-hair) theorem show that they are specified by only a small set of macroscopic parameters~\cite{Carter1971,Robinson1975}.

Beyond general relativity, modified gravity theories seek to account for galactic and extragalactic dynamics without invoking non-baryonic dark matter. A prominent example is Scalar–Tensor–Vector Gravity (STVG or ``MOG''), introduced by Moffat~\cite{Moffat2006JCAP}. In the MOG model, the effective gravitational coupling is dynamical, \(G = G_{N}(1+\alpha)\), and a massive vector field couples to matter through a gravitational charge \(Q_{g}=\sqrt{\alpha G_{N}}\,m\), leading to a Reissner–Nordstr\"{o}m-like \(+Q_{g}^{2}/r^{2}\) contribution to the metric function in the strong field  regime~\cite{Moffat2015EPJC}. The static Schwarzschild–MOG and rotating Kerr–MOG black hole solutions in this theory have been constructed, and their observational signatures on the black hole shadow have been explored~\cite{Moffat2015EPJC,Moffat2015Shadows}. On larger scales, the weak field limit of MOG has been applied to galaxy rotation curves and cluster dynamics with encouraging success~\cite{BrownsteinMoffat2006ApJ,Moffat2013MNRAS,Moffat2009MNRAS}. In addition, the geodesic motion~\cite{KerrMOG18,AccDiskImage} and non-geodesic motion~\cite{AccDiskMOG17,Turimov23,TurimovEPJP} of massive test particles in MOG background have been studied, and the accretion process into MOG black holes has been investigated~\cite{aJ19,kN23}.

A novel, complementary, fully relativistic parameter estimation program uses directly observable quantities from accretion disks to infer black hole parameters~\cite{ApJL,TXS,TenAGNs, FiveAGNs} and the Hubble constant~\cite{SdS}. In particular, the total frequency shift of photons emitted by test particles on circular equatorial orbits, together with a geometric aperture angle locating the emitter, encode the gravitational and kinematic information of the source along the line of sight~\cite{MCPII,MCPXI}. This, in turn, can be employed to extract the information of the central compact object by developing a suitable formalism~\cite{hanPRD2015,pBaHmMuNprd2022,KdS} (see also~\cite{dMmMaHepjc2024} for generic static spacetimes). The recently introduced \emph{redshift rapidity}, the derivative of the redshift with respect to the emitter’s proper time, acts as a relativistic invariant that helps resolve parameter degeneracies when expressions are evaluated on the circular and elliptical orbits~\cite{Momennia2024EPJC}. In the Schwarzschild background, this framework yields closed-form relations that disentangle the black hole mass and its distance from Earth using only directly measurable quantities~\cite{Momennia2024EPJC}. The method has been generalized to the Reissner–Nordstr\"{o}m spacetime, providing analytic formulas for the mass \(M\), electric charge \(Q\), and distance \(D\) in terms of observables~\cite{Morales2024RN}. More recently, it has been shown that the mass, rotation parameter, and distance to the Kerr black hole can be written in terms of fully observational quantities on the midline and close to the line-of-sight (LOS)~\cite{Momennia:2025:KerrParams}.

The methodology developed in~\cite{hanPRD2015,pBaHmMuNprd2022,KdS} has been employed to explore the properties of quantum-corrected black hole spacetimes by sending probe particles towards them~\cite{sH25}, and investigate the frequency shift of black holes in toroidal geometry~\cite{ToroidalBH}. Besides, the mass of a polymerized black hole and its related quantum parameter have been expressed in terms of a few direct observables~\cite{qFxZ23}, and a relation among the mass, spin, and charge parameters of a Kerr–Newman black hole with frequency-shifted photons has been obtained~\cite{rBetal25} (see also~\cite{nBaS26} and~\cite{bNaPjT25} for similar applications to the surface of compact stars and observing photons from an emitter falling into the Schwarzschild--de Sitter black hole, respectively). In addition, this formalism has been applied to Reissner–Nordstr\"{o}m black holes immersed in perfect fluid dark matter with the aim of deriving expressions for the total frequency shifts in terms of the black hole mass, electric charge, and the dark matter parameter~\cite{RNDM}. 

 Building on these developments, we adapt and extend the redshift/rapidity formalism to MOG spacetimes by considering massive geodesic emitters. For the spherically symmetric Schwarzschild–MOG geometry, we derive analytic, closed-form relations that express the black hole mass \(M\), the MOG coupling \(\alpha\), and the distance \(D\) to the observer in terms of the total frequency shift, the telescope’s aperture angle, and the redshift rapidity, evaluated on the midline for circular orbits in a thin accretion disk. On the other hand, we also obtain an expression for the distance to the black hole by incorporating the rapidity at the LOS. We then extend the construction to the rotating Kerr–MOG case by introducing the \emph{redshift acceleration}, a quantity that in the Newtonian limit corresponds to the classical jerk of the particle and that has been of increasing importance in astrophysics and cosmology \cite{Russo2009JPA, Ford2011Bayesian, Dutta2025AA, Martins2016PRD}. Similar to the static case, we obtain exact analytic formulas that determine the rotation parameter \(a\) jointly with \(M\), \(\alpha\), and \(D\) from the same class of directly measurable quantities on the midline. In the limit \(\alpha \to 0\), our expressions continuously reduce to their counterparts in the standard Schwarzschild and Kerr backgrounds~\cite{Momennia2024EPJC,Momennia:2025:KerrParams}, enabling a direct measurement of \(\alpha\) and thereby a quantitative test for deviations from general relativity. In the derived analytic closed-form formulas, the detector should be exactly in the emitter's plane and observe an edge-on view of the disk. For a general inclined disk, a more general inclined-observer formalism would be required. Note that an earlier study in the Kerr-MOG black hole spacetime was performed in~\cite{KerrMOG18} based on the kinematic redshift of massive geodesic emitters.

The Outline of this paper is as follows. Section~\ref{sec:geodesics} develops the geodesic motion of massive and null particles in a MOG background. We derive the nonvanishing components of the four–velocity for massive test particles, the photons’ four–momentum, and the associated impact parameter in the equator, all expressed in terms of the Schwarzschild–MOG black–hole mass \(M\), the MOG coupling \(\alpha\), and the orbital radius $r_e$. 
In Section~\ref{sec:freqshift}, we present our relativistic methodology in the MOG setting and, using the results of Section~\ref{sec:geodesics}, obtain the total frequency shift for emitters on circular equatorial orbits at arbitrary azimuth. We then provide midline and LOS formulas for the mass–to–radius ratio \(M/r_e\) and for \(\alpha\) in terms of directly observable quantities. 
Section~\ref{sec:rapidity} defines the \emph{redshift rapidity} as the proper–time derivative of the redshift in the MOG background, and combines the redshift and the redshift rapidity to obtain closed–form inversions for the black hole mass \(M\), the MOG parameter \(\alpha\), and the distance \(D\) solely in terms of observables. At the line-of-sight, we incorporated the redshift rapidity in the distance formula. Section~\ref{sec:kerrmog} extends the framework to the rotating Kerr–MOG geometry and uses the same observables and \emph{redshift acceleration} to infer the rotation parameter \(a\) together with \(M\), \(\alpha\), and \(D\). Finally, Section~\ref{sec:conclusion} summarizes our results, discusses their observational applications, and outlines their broader relevance.

\section{GEODESIC MOTION IN THE SCHWARZSCHILD--MOG SPACETIME}
\label{sec:geodesics}

In this section, we establish the relativistic framework and derive the nonvanishing components of the four–velocity for a test particle orbiting a MOG black hole by analyzing its equations of motion in terms of the underlying metric functions. We likewise obtain the four–momentum of null geodesics and express the corresponding impact parameter directly in terms of the Schwarzschild–MOG parameters \(M\) and \(\alpha\). These kinematic ingredients will be used in the next section to construct the total frequency–shift formula in the MOG background.

We consider the line element around a static, spherically symmetric Schwarzschild–MOG black hole, written in Schwarzschild coordinates $\left( t,r,\vartheta ,\varphi \right)$ as
[we use $c=1=G_{N}=k_c$ units]: 
\begin{equation}
ds^{2}=-g(r)dt^{2}+g^{-1}(r)dr^{2}+r^{2}d\vartheta ^{2}+r^2 \sin ^{2}\vartheta d\varphi ^{2},  \label{metric1}
\end{equation}
in which the metric function  $g(r)$ is given by \cite{Moffat2015EPJC}
\begin{equation}
 g(r)= 1-\frac{2(1+ \alpha)M}{r}+\frac{\alpha(1+\alpha) M^2}{r^2},
\label{g}
\end{equation}
where $M$ is the black hole mass and $\alpha$ is the MOG parameter. 

In the Schwarzschild--MOG spacetime, the physical curvature singularity lies at the origin \(r=0\). The two coordinate singularities are determined by the roots of the metric function \(g(r)=0\) and identify, respectively, the Cauchy (inner) and event (outer) horizons. Solving \(g(r)=0\) yields the horizon radii \cite{Moffat2015EPJC}
\begin{equation}
 r_{\pm} = M \left[1 + \alpha \pm (1+\alpha )^\frac{1}{2} \right],
 \label{Horizons}
\end{equation}
where reduces to the Schwarzschild radius $r_+=2M$ in the limit $\alpha \to 0$. Generally, the non-vanishing MOG parameter $\alpha$ exhibits deviations from the standard Schwarzschild solutions. In this study, we aim to find closed formulas for $\alpha$ in terms of directly measurable elements, thereby enabling the measurement of deviations from general relativity.

Test particles and photons in the vicinity of a modified Schwarzschild black hole respond to the spacetime curvature sourced by the mass \(M\) and the MOG coupling \(\alpha\). The corresponding kinematic and optical observables, in particular the frequency shift of photons emitted by orbiting matter, encode information about the spacetime geometry and these parameters. Consequently, it is necessary to analyze the motion of massive and null geodesics in the Schwarzschild--MOG background defined by Eq.~(\ref{metric1}); this is the focus of the following subsections.

\subsection{Geodesics of massive particles}
\label{nonnull}

In this section, we analyze the geodesic motion of massive test particles revolving around a Schwarzschild–MOG black hole [see Appendix~\ref{NONgeodesic} for a short discussion on the non-geodesic motion]. A neutral massive test particle moving along geodesics of the Schwarzschild--MOG spacetime has four–velocity
\begin{equation}
U^\mu = (U^t, U^r, U^\vartheta, U^\varphi), \quad U^{\mu}=\frac{dx^\mu}{d\tau},
\end{equation}
subject to the normalization
\begin{equation}
U^\mu U_\mu = -1,
\label{normal}
\end{equation}
where $\tau$ is the proper time.

To model thin accretion disks around supermassive black holes, we confine the motion to a plane. Because the metric~(\ref{metric1}) is spherically symmetric, we may choose the equatorial plane, $\vartheta=\pi/2$, without loss of generality. In this case $U^\vartheta=0$ and the $g_{\varphi \varphi}$ component of the metric simplifies to $r^2$. The timelike and axial Killing vector fields of the MOG metric, $\xi^{\mu}=\delta_t^\mu$ and $\psi^{\mu}=\delta_\varphi^\mu$, furnish the conserved energy and angular momentum per unit rest mass:
\begin{equation}
L=\frac{\bar{L}}{m}=\psi _{\mu}U^{\mu}=g_{\mu \nu}\psi ^{\nu}U^{\mu}=r^2 U^{\varphi},
    \label{killin2}
\end{equation}
\begin{equation}
    E=\frac{\bar{E}}{m}=-\xi _{\mu}U^{\mu}=-g_{\mu \nu}\xi ^{\nu}U^{\mu}= g(r)U^t.
    \label{killn}
\end{equation}

Inserting $U^{t}$ and $U^{\varphi}$ from (\ref{killin2})-(\ref{killn}) into the normalization condition (\ref{normal}) gives
\begin{equation}
(U^r)^2+g(r)+\frac{g(r) }{r^2}L^2=E^2.
\label{veff}
\end{equation}
This has the structure of an energy balance for a nonrelativistic particle moving in the effective potential
\begin{equation}
V_{eff}=g(r) \left(1+\frac{L^2}{r^2}\right).
\label{Veff2}
\end{equation}

We now consider the \emph{special case} of circular motion, for which the radial component in (\ref{veff}) satisfies $U^r=0$. Consequently, the conditions $V_{eff}=E^2$ and $V'_{eff}=0$ characterize circular orbits leading to:
\begin{equation}
g(r) \left(1+\frac{L^2}{r^2}\right)=E^2,
   \label{vefd}
\end{equation}
\begin{equation}
V'_{eff}(r)=g'(r)\left(1+\frac{L^2}{r^2}\right)-g(r)\frac{2L^2}{r^3}=0,
\label{veffprima}
\end{equation}
where prime refers to derivative with respect to the radial coordinate $r$. Now, one can find the explicit form of the total energy and angular momentum of massive test particles in terms of spacetime parameters $M$ and $\alpha$ by solving Eqs. (\ref{vefd}) and (\ref{veffprima}) as follows
\begin{equation}
\left. E=g(r)\sqrt{\frac{2}{2g(r)-rg'(r)}}\right|_{r=r_e} ,
\label{energy}
\end{equation}
\begin{equation}
    \left. L =(\pm)r\sqrt{\frac{rg'(r)}{2g(r)-rg'(r)}} \right|_{r=r_e},
    \label{energy2}
\end{equation}
where $r_e$ denotes the emitter’s orbital radius and $g(r)$ is the metric function given in Eq. (\ref{g}). In what follows, we focus on the clockwise rotation of the emitter only (plus sign of $L$) and discard counterclockwise motion without loss of generality.

Next, using Eqs.~(\ref{g}), (\ref{killin2}), (\ref{killn}), (\ref{energy}), and (\ref{energy2}), the nonvanishing temporal and azimuthal components of the emitter’s four–velocity can be expressed in terms of the black hole mass $M$ and MOG parameter $\alpha$:
\begin{equation}\label{Ut}
   U^t_e=\frac{E}{g(r_e)}=\frac{r_e}{\sqrt{r_e^2-3(1+\alpha)Mr_e+2\alpha(1+\alpha)M^2}}  , 
\end{equation}
\begin{equation}\label{Uphi}
     U^{\varphi}_e=\frac{L}{r_e^2}=\frac{1}{r_e}\sqrt{\frac{(1+\alpha)Mr_e-\alpha(1+\alpha)M^2}{r_e^2-3(1 + \alpha)Mr_e+2\alpha(1+\alpha) M^2}}.
\end{equation}
This correspondence between $\{M,\alpha\}$ and the four–velocity components will be essential in what follows, as it allows us to extract the information about the spacetime curvature encoded in the observed frequency shifts in the MOG background.

Finally, orbital stability is assessed via the second derivative of the radial effective potential in (\ref{Veff2}). Stability requires $V''_{eff} \ge 0$, while $V''_{eff}=0$ identifies the innermost stable circular orbit (ISCO). The second derivative reads
\begin{align}\label{estable}
V''_{eff}&= g''(r)\left(1+\frac{L^2}{r^2}\right)-g'(r)\frac{4L^2}{r^3}+g(r)\frac{6L^2}{r^4}.
\end{align}
Substituting $g(r)$ and $L$ from Eqs.~(\ref{g}) and (\ref{energy2}) into (\ref{estable}) and imposing $V''_{eff}=0$ leads to the cubic equation
\begin{equation}
\begin{aligned}
 r^3 &- 6(1+\alpha) M r^2 + 9\alpha(1+\alpha) M^2 r
- 4\alpha^2 (1+\alpha) M^3 = 0.
\end{aligned}
\end{equation}
whose physically relevant real root is
\begin{eqnarray}
r_{ISCO}&=&(\alpha+1) M \left\{\left[\frac{ \alpha ^2+\left(\sqrt{\alpha +5}+7\right) \alpha +8}{(\alpha +1)^2}\right]^{1/3} \right. \notag \\
&&\left. + \frac{(\alpha +4)(\alpha +1)^{-1/3}}{ \left[ \alpha ^2+\left(\sqrt{\alpha +5}+7\right) \alpha +8 \right]^{1/3} }+2 \right\}.
\label{Risco}
\end{eqnarray}

This radius marks the boundary of stability and the inner edge of the thin stable accretion disk circularly orbiting the Schwarzschild-MOG black hole. It is worth mentioning that $r_{ISCO}$ is always larger than the corresponding ISCO radius in the standard Schwarzschild black hole background for nonvanishing MOG parameter $\alpha$ [see Lemma \ref{IscoLem} below]. Throughout, we consider emitters on circular orbits satisfying $r_e \ge r_{\mathrm{ISCO}}$.

\begin{lemma}
\label{IscoLem} Let $M>0$ and $\alpha> 0$. In addition, let $r_{ISCO}^{(MOG)}$ be the ISCO radius in the Schwarzschild-MOG spacetime given by Eq. (\ref{Risco}) and $r_{ISCO}^{(Schw)}=6M$ be the ISCO radius in the standard Schwarzschild background, such that $r_{ISCO}^{(MOG)} (\alpha=0) = r_{ISCO}^{(Schw)}$. Then ISCO radius in Schwarzschild-MOG spacetime is larger than ISCO radius in the standard Schwarzschild background: $r_{ISCO}^{(MOG)} > r_{ISCO}^{(Schw)}$.
\end{lemma}

\proofof{Lemma~\ref{IscoLem}} In order to prove this lemma, we first rewrite the ISCO radius (\ref{Risco}) as 
\begin{eqnarray}
\frac{r_{ISCO}^{(MOG)}}{M}&=&x_F+x_S+x_T,
\label{DecomISCO}
\end{eqnarray}
with
\begin{eqnarray}
x_F&=&2 (\alpha +1), \label{xF}\\
x_S&=&(\alpha +1)^{1/3}\left[\alpha ^2+\left(\sqrt{\alpha +5}+7\right) \alpha +8\right]^{1/3}, \label{xS}\\
x_T&=&\frac{(\alpha +4) (\alpha +1)^{2/3}}{\left[\alpha ^2+\left(\sqrt{\alpha +5}+7\right) \alpha +8\right]^{1/3}}  \label{xT}.
\end{eqnarray}
From Eqs. (\ref{xF})-(\ref{xS}), it is obvious that $x_F>2$ and $x_S>2$ for $\alpha>0$. Besides, from Eq. (\ref{xT}), we find that $x_T$ is positive-definite, $x_T(\alpha=0)=2$, and it scales as $x_T(\alpha) \approx \alpha$ in the limit $\alpha \to \infty$. In addition, from the derivative of $x_T$ with respect to $\alpha$
\begin{eqnarray}
\frac{dx_T}{d\alpha}&=&\frac{\left[\alpha  \left(\alpha +\sqrt{\alpha +5}+7\right)+8 \right]^{-4/3}}{6(\alpha +1)^{1/3}\sqrt{\alpha +5}}\times \notag\\
&&\left[\left(6 \sqrt{\alpha +5}+7\right) \alpha ^3+\left(58 \sqrt{\alpha +5}+47\right) \alpha ^2 \right. \notag\\
&&\left. +4 \left(37 \sqrt{\alpha +5}+12\right) \alpha +40 \left(3 \sqrt{\alpha +5}-1\right)\right], \qquad
\end{eqnarray}
one sees that $dx_T/d\alpha>0$ since $3 \sqrt{\alpha +5}\geq  3\sqrt{5} >1$. From these observations, it follows that $x_T$ is a smooth and monotonically increasing function of $\alpha$ such that $x_T > 2$ for $\alpha>0$. Therefore, since $x_F,x_S,x_T>2$ for $\alpha>0$, from Eq. (\ref{DecomISCO}) we have
\begin{eqnarray}
\frac{r_{ISCO}^{(MOG)}}{M}>\frac{r_{ISCO}^{(Schw)}}{M}, \qquad \frac{r_{ISCO}^{(Schw)}}{M}=6,
\end{eqnarray}
\hfill \fbox{} 
\vspace{0.3cm}

\subsection{Geodesics of null particles in a MOG spacetime}

We now examine the trajectories of photons emitted by the massive particles discussed in the previous section. The photons follow null geodesics with four–momentum $k^{\mu}$ obeying
\begin{equation}
    k^{\mu}k_{\mu}=0.
    \label{momento}
\end{equation}

Because the spacetime is spherically symmetric, the motion admits conserved quantities: the photon energy $E_{\gamma}$ and angular momentum $L_{\gamma}$, given by
\begin{equation}
     E_{\gamma}=-\xi _{\mu}k^{\mu}=-g_{\mu \nu}\xi ^{\nu}k^{\mu}= g(r)k^t, 
\end{equation}
\begin{equation}
     L_{\gamma}=\psi _{\mu}k^{\mu}=g_{\mu \nu}\psi ^{\nu}k^{\mu}=r^2 \sin^2 \vartheta k^{\varphi},
\end{equation}
where the subscript $\gamma$ labels photonic quantities. From these definitions, the temporal and azimuthal components of the four–momentum can be written in terms of $E_{\gamma}$ and $L_{\gamma}$ as
\begin{equation}
    \left. k^t_e= \frac{E_\gamma}{g(r) }\right|_{r=r_e}=\frac{E_{\gamma}}{g(r_e)},
\label{energyphoton}
\end{equation}
\begin{equation}
     \left. k^{\varphi}_e=\frac{L_{\gamma}}{g_{\varphi \varphi}}\right|_{r=r_e}=\frac{L_{\gamma}}{r_e^2\sin ^{2}\vartheta},
     \label{eneryphoton2}
\end{equation}
where they are evaluated at the emission point. Substituting (\ref{energyphoton}) and (\ref{eneryphoton2}) into the null condition (\ref{momento}) yields
\begin{equation}
(k^r )^2 _e=E_{\gamma}^2-g(r_e)\frac{L_\gamma^2}{r_e^2},
\label{4momento}
\end{equation}
for the radial component where we restricted the motion of the photons to the equatorial plane ($k^\vartheta= 0$).

In terms of the MOG black hole parameters, the impact parameter reads~\cite{Momennia2024EPJC}
\begin{equation}\label{lbp}
     b_{\varphi}=\frac{L_\gamma}{E_\gamma}=-\frac{r_e\sin(\varphi+\delta)}{\sqrt{g(r_e)\sin^2(\varphi+\delta)+\cos^2(\varphi+\delta)}},
\end{equation}
that captures the light bending experienced by photons emitted from any position along a circular emitter orbit. In this relation, $\varphi$ is the azimuthal angle and $\delta$
is the aperture angle of the telescope.

The impact parameter $b_{\varphi}$ encodes the deflection produced by the gravitational field of the MOG black hole (characterized by $M$ and $\alpha$) and remains constant along the photon’s null geodesic, from emission to detection. Since $E_{\gamma}$ and $L_{\gamma}$ are conserved, we have $b_{\varphi,e}=b_{\varphi,d}$, with the subscripts denoting emission and detection points, respectively.

\begin{figure*}[t]
\centering
\includegraphics[scale=.9]{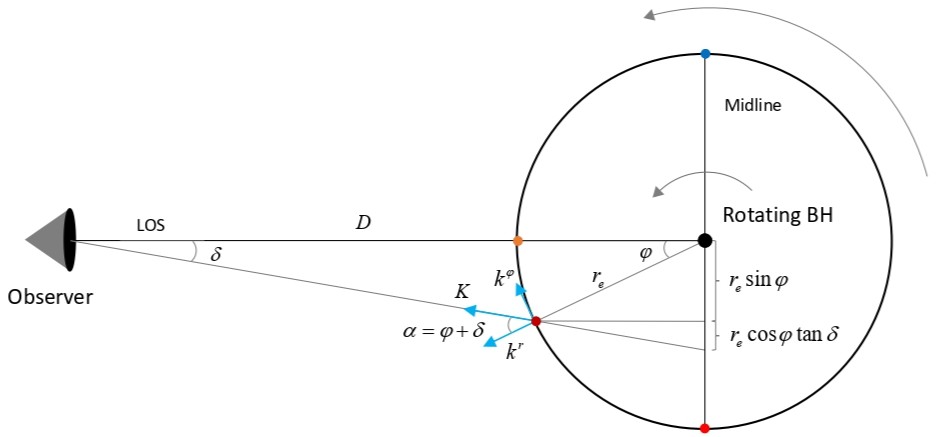}
\caption{The schematic illustration of a co-rotating emitter circularly orbiting a rotating MOG black hole at the equator. This figure shows the relationship between the azimuthal angle $\varphi$ and the angular distance $\delta$ (this figure is taken from Ref.~\cite{Momennia:2025:KerrParams}).}
\label{PhiDeltaFig}
\end{figure*}

Finally, the angles $\varphi$ and $\delta$ satisfy the geometric relation~\cite{Momennia2024EPJC}
\begin{equation}\label{rel2}
D\sin{\delta}=r_e\sin(\varphi+\delta),
\end{equation}
where $D$ is the distance between the black hole center and the distant observer (see Fig.~\ref{PhiDeltaFig} for a schematic illustration of the relationship between the observable angle $\delta$ and the unobservable angle $\varphi$).

Solving (\ref{rel2}) allows one to express $\delta$ in terms of the remaining parameters. In general there are four branches and the physical choice used here is
\begin{equation}\label{deltaphirel}
    \delta(\varphi)=\arccos{\left(\frac{D-r_e\cos{\varphi}}{\sqrt{D^2+r^2_e-2Dr_e\cos{\varphi}}}\right)}.
\end{equation}

\section{Frequency shift in THE SCHWARZSCHILD--MOG SPACETIME}\label{sec:freqshift}

After electromagnetic emission by massive test particles orbiting the Schwarzschild-MOG black hole, photons propagate through the black hole’s gravitational field; consequently, the detected frequency by a distant observer differs from that at emission. The photon frequency at a spacetime point $x_{p}^{\mu}$ is
\begin{equation}
\omega _{p}=-\left( k_{\mu }U^{\mu }\right) \big|_{p}\,,  \label{freq}
\end{equation}
where the index $p$ denotes either the emission event $x_{e}^{\mu }$ or the detection event $x_{d}^{\mu }$.

For static, spherically symmetric metrics of the form (\ref{metric1}), the frequency shift takes the form~\cite{hanPRD2015}
\begin{eqnarray}
1 &+&z_{_{BH}}\!=\frac{\omega _{e}}{\omega _{d}}  \notag \\
&=&\frac{(E_{\gamma }U^{t}-L_{\gamma }U^{\varphi}
-g^{-1}(r) U^{r}k^{r}-g_{\vartheta \vartheta }U^{\vartheta }k^{\vartheta })\big| _{e}}{%
(E_{\gamma }U^{t}-L_{\gamma }U^{\varphi }-g^{-1}(r)U^{r}k^{r}-g_{\vartheta \vartheta
}U^{\vartheta }k^{\vartheta })\big| _{d}}\,.  \qquad \label{GeneralShift}
\end{eqnarray}

In the modified gravity setting, Eq.~(\ref{GeneralShift}) is specified by the photon four–momentum and the emitter’s four–velocity given in Eqs.~(\ref{Ut})–(\ref{Uphi}), (\ref{energyphoton})–(\ref{eneryphoton2}), and (\ref{4momento}). We also recall that the radial and polar components vanish for both emitter and detector, $U_p^r=U_p^{\vartheta}=0$. Placing the detector at a large distance $r_d=D\gg r_e$ further implies
\begin{equation}
U_d^{\mu} = \delta^\mu_t,
\label{Detector4Velocity}
\end{equation}
since $U_d^{\varphi}\to 0$ and $U_d^{t}\to 1$ as $r_d\to\infty$ (see Eqs. (\ref{Ut})–(\ref{Uphi}) while replacing subscript ``$e$" with ``$d$"), hence the observer is effectively outside the black hole's gravitational field. Under these conditions, Eq.~(\ref{GeneralShift}) simplifies to
\begin{align} \label{Redshift}
    1+z_{MOG} &=\frac{E_{\gamma}U^t_e-L_{\gamma}U^{\varphi}_e}{E_{\gamma}}=U^t_e-b_{\varphi}U^{\varphi}_e\notag\\
    &=\frac{1}{\sqrt{1-3\tilde{P}+2\tilde{N}}}\times \notag\\
    &{\left[1+\frac{\sqrt{\tilde{P}-\tilde{N}}\sin(\varphi+\delta)}{\sqrt{\cos^2(\varphi+\delta)+g(r_e)\sin^2(\varphi+\delta)}}\right],}
\end{align}
which is the frequency–shift relation in the Schwarzschild MOG spacetime and, as expected, reduces to the Schwarzschild limit when $\alpha=0$ \cite{Momennia2024EPJC}. Here we used the shorthand notation $\tilde{P}=(1+ \alpha)M/r_e$ and $\tilde{N}=\alpha(1+\alpha)M^2/r_e^{2}$, and Eqs.~(\ref{Ut})–(\ref{Uphi}) together with (\ref{lbp}) have been used in the last step. In Eq.~(\ref{Redshift}), $z_{MOG}$ and $\delta$ are directly observable, whereas $r_e$ and $\varphi$ are not; the quantities to be inferred are $M$ and $\alpha$. It is worth mentioning that since the structure of the metric function~(\ref{g}) is quite similar to Reissner-Nordstr\"{o}m solutions, one expects to observe similar behavior in the frequency shift relation~(\ref{Redshift}). However, interestingly, unlike the Reissner-Nordstr\"{o}m black hole~\cite{Morales2024RN}, the absolute value of $z_{MOG}$ is \emph{always} larger than the
corresponding total frequency shift in the standard Schwarzschild spacetime at the midline $\varphi=\pm \pi/2$ [see Proposition \ref{Pro} below].

Moreover, Eq.~(\ref{Redshift}) separates the gravitational and kinematic pieces for circular motion around a Schwarzschild MOG black hole:
\begin{equation}
    z_g = U_e^t - 1,
    \label{GravRed}
\end{equation}
\begin{equation}
    z_{kin}=-b_{\varphi}U_e^{\varphi}.
    \label{KinRed}
\end{equation}

In the following proposition, we prove that $z_g$ is always larger than the
corresponding gravitational redshift in the standard Schwarzschild
black hole background for nonvanishing MOG parameter $\alpha$. We also prove that $|z_{MOG}|$ and $|z_{kin}|$ are always larger than the
corresponding total and kinematic frequency shifts in the standard Schwarzschild spacetime for nonzero values of $\alpha$ at the midline $\varphi=\pm \pi/2$.

\begin{proposition}
\label{Pro} Let $M>0$, $\alpha> 0$, $r_e>r_{ISCO}$, and $\varphi=\varphi_m=\pm \pi/2$. In addition, let $z_{MOG}$ be the total frequency shift in the Schwarzschild-MOG spacetime given by Eq. (\ref{Redshift}) and $z_{Schw}$ be the total frequency shift in the standard Schwarzschild background, such that $z_{MOG}(\alpha=0) = z_{Schw}$. Then the absolute total frequency shift in the Schwarzschild-MOG spacetime is larger than the absolute total frequency shift in the standard Schwarzschild background: $\left|
z_{MOG}\right| > \left|z_{Schw}\right|$.
\end{proposition}

In order to prove this proposition, we first need to show that the gravitational redshift and absolute kinematic redshift at the midline in the Schwarzschild-MOG spacetime are larger than the corresponding ones in the standard Schwarzschild background. We show these through Lemma \ref{ZgLem} and Lemma \ref{ZkinLem} below:

\begin{lemma}
\label{ZgLem} Let $M>0$, $\alpha> 0$, and $r_e>r_{ISCO}$. In addition, let $z_g^{(MOG)}$ be the gravitational redshift in the Schwarzschild-MOG spacetime given by Eq. (\ref{GravRed}) and $z_g^{(Schw)}$ be the gravitational redshift in the standard Schwarzschild background, such that $z_g^{(MOG)}(\alpha=0) = z_g^{(Schw)}$. Then $z_g^{(MOG)}> z_g^{(Schw)}$.
\end{lemma}

\proofof{Lemma~\ref{ZgLem}} First from Eq. (\ref{Horizons}) note that $r_e>r_+=M(1+\alpha+\sqrt{1+\alpha})$. This leads to 
\begin{eqnarray}
    (1+\alpha)\frac{M}{r_e}<1.
    \label{CondMr}
\end{eqnarray}
After some manipulation, from this inequality, we find that
\begin{eqnarray}
    -3\alpha\frac{M}{r_e}+2\alpha (1+\alpha) \frac{M^2}{r_e^2}<0.
\end{eqnarray}
Finally, we add $1-3M/r_e$ to both sides, invert the result and take the square root to get
\begin{eqnarray}
    \frac{1}{\sqrt{1-3(1+\alpha)\frac{M}{r_e}+2\alpha (1+\alpha) \frac{M^2}{r_e^2}}}> \frac{1}{\sqrt{1-3\frac{M}{r_e}}}.
\end{eqnarray}
which from this it follows that $z_g^{(MOG)}> z_g^{(Schw)}$. Because this inequality is valid for $r_e>r_+$, it also holds for $r_e \geq r_{ISCO}>r_+$.

\hfill \fbox{} 
\vspace{0.3cm}

\begin{lemma}
\label{ZkinLem} Let $M>0$, $\alpha> 0$, $r_e>r_{ISCO}$, and $\varphi=\varphi_m=\pm \pi/2$. In addition, let $z_{kin}^{(MOG)}$ be the kinetic redshift in the Schwarzschild-MOG spacetime given by Eq. (\ref{KinRed}) and $z_{kin}^{(Schw)}$ be the kinetic redshift in the standard Schwarzschild background, such that $z_{kin}^{(MOG)}(\alpha=0) = z_{kin}^{(Schw)}$. Then $\left|z_{kin}^{(MOG)}\right| > \left|z_{kin}^{(Schw)}\right|$.
\end{lemma}

\proofof{Lemma~\ref{ZkinLem}} After some manipulation, from the inequality (\ref{CondMr}), we find that
\begin{eqnarray}
    \alpha\frac{M}{r_e}-\alpha (1+\alpha) \frac{M^2}{r_e^2}>0.
\end{eqnarray}
Now, we add $M/r_e$ to both sides and take the square root to get
\begin{eqnarray}
    \sqrt{(1+\alpha)\frac{M}{r_e}-\alpha (1+\alpha) \frac{M^2}{r_e^2}} |\cos \delta|> \sqrt{\frac{M}{r_e}}|\cos \delta|.
    \label{Numerator}
\end{eqnarray}

On the other hand, Eq. (\ref{CondMr}) also gives
\begin{eqnarray}
    -2\alpha\frac{M}{r_e}+\alpha (1+\alpha) \frac{M^2}{r_e^2}<0.
\end{eqnarray}
Then, after some manipulation including adding $1-2M/r_e$ to both sides, invert the result, and take the square root, it leads to
\begin{eqnarray}
    \frac{1}{\sqrt{\sin^2 \delta+\left[1-2(1+\alpha)\frac{M}{r_e}+\alpha (1+\alpha) \frac{M^2}{r_e^2}\right]\cos^2 \delta}}> \notag \\
    \frac{1}{\sqrt{\sin^2 \delta+\left[1-2\frac{M}{r_e}\right]\cos^2 \delta}},\quad
\end{eqnarray}
which means 
\begin{eqnarray}
    \frac{1}{\sqrt{\sin^2 \delta+g(r_e) \cos^2 \delta}}> \frac{1}{\sqrt{\sin^2 \delta+g_{Schw}(r_e)\cos^2 \delta}},\quad
    \label{Denominator}
\end{eqnarray}
where $g_{Schw}(r_e)=g(r_e,\alpha=0)$ is the Schwarzschild metric function.

Finally, Eqs. (\ref{Numerator}) and (\ref{Denominator}) lead to
\begin{eqnarray}
     \sqrt{\frac{\tilde{P}-\tilde{N}}{\sin^2 \delta+g(r_e) \cos^2 \delta}} |\cos \delta| > \notag \\ \sqrt{\frac{\frac {M} {r_e}}{\sin^2 \delta+g_{Schw}(r_e)\cos^2 \delta}} |\cos \delta|.\quad
\end{eqnarray}
where by considering Eqs. (\ref{Redshift}) and (\ref{KinRed}), we find that $\left|z_{kin}^{(MOG)}\right| > \left|z_{kin}^{(Schw)}\right|$.

\hfill \fbox{} 
\vspace{0.3cm}

Since $z_{MOG}=z_{g}^{(MOG)}+z_{kin}^{(MOG)}$ and $z_{Schw}=z_{g}^{(Schw)}+z_{kin}^{(Schw)}$, it follows from Lemma \ref{ZgLem} and Lemma \ref{ZkinLem} that $\left|
z_{MOG}\right| > \left|z_{Schw}\right|$.

\begin{figure*}[t]
\centering
\includegraphics[width=\textwidth]{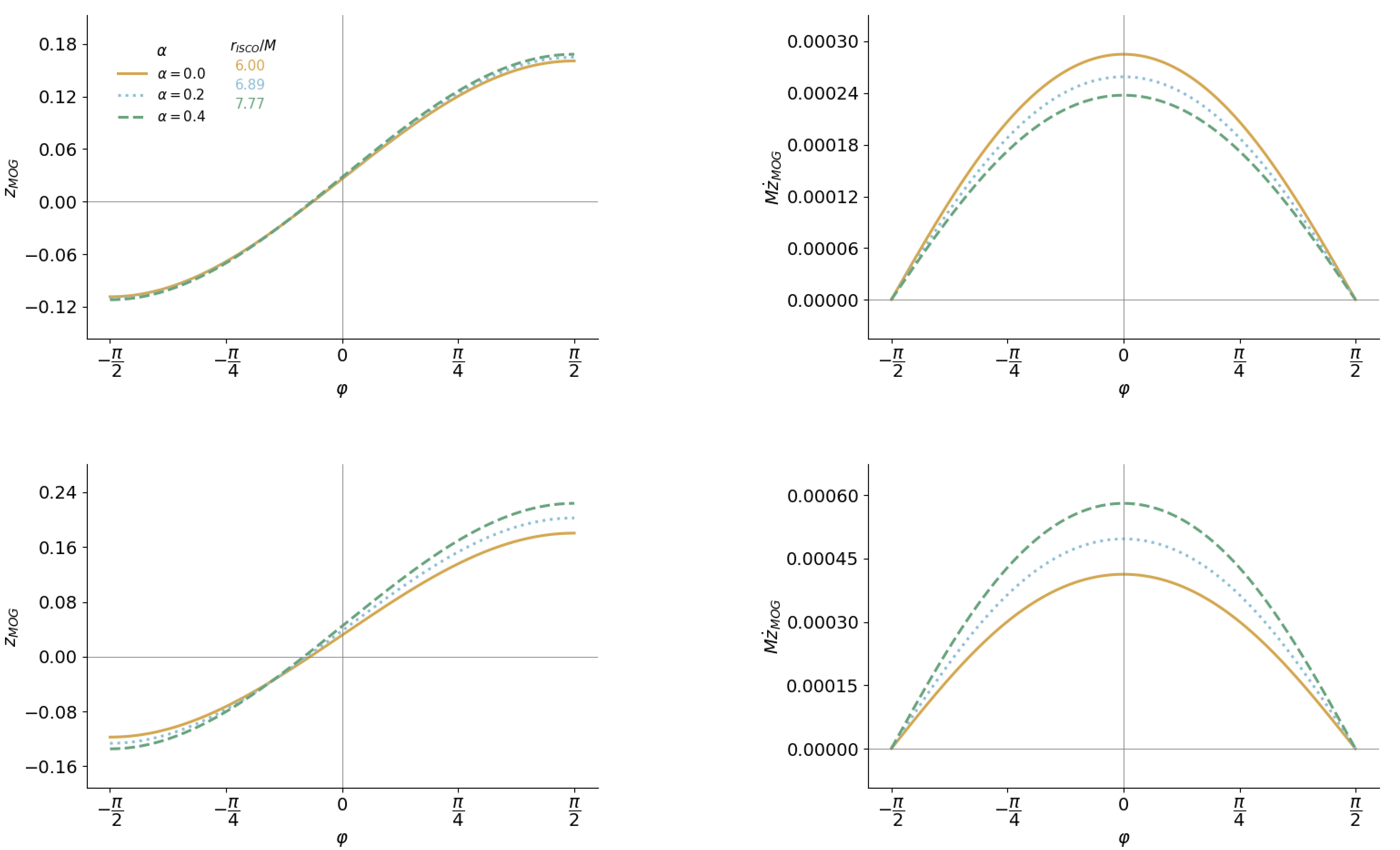}
\caption{The frequency shift $z_{MOG}$ and the redshift rapidity $\dot{z}_{MOG}$ versus the azimuthal angle in the MOG background for $r_{e}=10r_{ISCO}$, $D=10^5r_{ISCO}$ (upper panels) and $r_{e}=50M$, $D=10^5M$ (lower panels), and different values of the MOG parameter $\alpha$. The redshift and blueshift are largest on the midline, $\varphi\approx\pm \pi/2$, while the redshift rapidity peaks along the line of sight, $\varphi=0$. The continuous orange curves show the frequency shift and redshift rapidity in the standard Schwarzschild spacetime. To produce these curves, we substitute $r_{ISCO}$ (upper panels) and $\delta$ (all panels) from Eqs.~(\ref{Risco}) and (\ref{deltaphirel}) into Eqs.~(\ref{Redshift}) and (\ref{RRgeneral}).}
\label{zDotFig}
\end{figure*}

The left panels of Fig.~\ref{zDotFig} display Eq.~(\ref{Redshift}) as a function of the azimuthal angle $\varphi$ for different values of the MOG parameter $\alpha$. In the upper-left panel we set $r_e=10\,r_{ISCO}(\alpha)$ and $D=10^{5}r_{ISCO}(\alpha)$, so the emitter radius grows with increasing $\alpha$ because $r_{ISCO}$ increases. As a consequence, the overall magnitude of the frequency shift increases slightly as $\alpha$ increases. In the lower-left panel we instead fix the emitter location and the observer distance to $r_e=50M$ and $D=10^{5}M$ for all black holes; in this case the shift increases with $\alpha$ since the azimuthal component of the 4-velocity is an increasing function of $\alpha$. In both cases, the total frequency shift is largest on the midline ($\varphi\approx\pm \pi/2$), which facilitates observational identification. It is worth noting that in consistency with the Proposition \ref{Pro}, Fig.~\ref{zDotFig} shows that the modifications due to the modified gravity parameter $\alpha$ lead to a higher frequency shift on the midline compared to the standard Schwarzschild black hole (denoted by continuous curves), indicating deviations from the general theory of relativity. In addition, at the LOS where $\varphi=0=\delta$, $z_g$ is always greater than the gravitational redshift in the standard Schwarzschild background as expected [see Lemma \ref{ZgLem}].

Finally, one may check that in the Newtonian
(weak--field) regime characterized by $M/r_e \to 0$, the Schwarzschild--MOG
redshift expression (\ref{Redshift}) reduces at leading order to
\begin{equation}
z_{\text{Newton}}=\sqrt{(1+\alpha) \frac{M}{r_e}}\sin(\varphi+\delta),
\label{RedshiftWeakField}
\end{equation}
where $\alpha$ shows MOG correction to Newtonian gravity. In the absence of modified gravity correction, $\alpha=0$, this leading contribution is simply the Doppler shift associated with the Keplerian
orbital speed of a particle on a circular orbit projected along the line of sight, as expected.

It is important to clarify that the parameter $\alpha$ appearing in
Eq.~\eqref{RedshiftWeakField} should be interpreted as the effective large-scale
MOG parameter relevant to galaxy rotation curves, galactic clusters, and black hole phenomenology.
In the full STVG/MOG weak-field regime, the dynamics is not controlled by
this enhanced coupling alone: the massive vector field also contributes to
the test-particle acceleration on small scales and cannot be neglected in Solar system
tests. According to the standard STVG/MOG weak-field treatment~\cite{Moffat2013MNRAS,MoffatToth2010,Moffat2014,MoffatToth2009},
the vector-field contribution suppresses the local enhancement associated with
$\alpha$, leading effectively to $\widetilde G\simeq G_N$ in the Solar system regime. On the other hand, on galactic and
cluster scales, the screening ceases to be operative and we have
$\widetilde G\simeq G_N(1+\alpha)$. Thus, the weak-field expression
\eqref{RedshiftWeakField} is to be understood as the effective large-scale limit of
the MOG model [see Appendix~\ref{NONgeodesic} for screened $\alpha$ in the weak-field limit due to the vector-field contribution].

\subsection{Redshift at the midline}

To write the mass–to–radius ratio and the MOG parameter in terms of observables, we evaluate the redshift and blueshift when the emitters lie on the midline $\varphi=\pm \pi /2$. At these azimuths, the impact parameter is extremal and Eq.~(\ref{Redshift}) becomes

\begin{align}\label{blueandred}
    1+z^{(m)}_{MOG_{1,2}} &= \frac{1}{\sqrt{1 - 3\tilde{P} + 2\tilde{N}}} \times \notag\\
    &\Biggl[\,1 \pm 
    \frac{\sqrt{\tilde{P} - \tilde{N}}\cos{\delta_m}}%
    {\sqrt{\sin^2{\delta_m} + (1 - 2\tilde{P} + \tilde{N})\cos^2{\delta_m}}}\,\Biggr],
\end{align}
where the plus sign corresponds to the receding (redshifted) side ($z^{(m)}_{MOG_1}$) and the minus sign to the approaching (blueshifted) side ($z^{(m)}_{MOG_2}$). Besides, here and in what follows, the index ``$m$'' refers to the observational parameters which should be measured at the midline.

Calling $ w_m   := (2 +z^{(m)}_{MOG_1} + z^{(m)}_{MOG_2})^{2}$ and $ h_m := (1+z^{(m)}_{MOG_1})(1+{z^{(m)}_{MOG_2}})$, from Eq.~(\ref{blueandred}) we obtain the system
\begin{equation}\label{rb1}
 h_m=\frac{1}{1-3\tilde{P}+2\tilde{N}}\left[\frac{\sec^2{\delta_m}-3\tilde{P}+2\tilde{N}}{\sec^2{\delta_m}-2\tilde{P}+\tilde{N}} \right],
\end{equation}
\begin{equation}\label{rb2}
    w_m=\frac{4}{1-3\tilde{P}+2\tilde{N}}.
\end{equation}

Note that $z^{(m)}_{MOG_{1,2}}$ are observational quantities; hence, in what follows, we call $w_m$ and $h_m$ as observational elements instead of $z^{(m)}_{MOG_{1,2}}$ for the sake of simplicity. Inverting (\ref{rb1})-(\ref{rb2}) yields
\begin{equation}
\tilde{P}=\frac{4}{w_m}-\frac{4+w_m\tan^2{\delta_m}}{2h_m}+2\sec^2{\delta_m}-1,
\label{tildeM}    
\end{equation}

\begin{equation}
\tilde{N} = \frac{8}{w_m}-\frac{12+3w_m\tan^2{\delta_m}}{4h_m}+3\sec^2{\delta_m}-2.
\label{tildeQ}
\end{equation}

Recalling the definition of $\tilde{P}$ and $\tilde{N}$

\begin{align}\label{m}
    \frac{(1+\alpha)M}{r_{e}} =   \tilde{P}, \qquad  \frac{\alpha(1+\alpha)M^2}{r_e^2} = \tilde{N},
\end{align}
whose solution for $\alpha$ and $M/r_e$ is
\begin{align}\label{m}
    \alpha = \frac{\tilde{N}}{\tilde{P}^2-\tilde{N}}, \qquad \frac{M}{r_e} =\tilde{P}-\frac{\tilde{N}}{\tilde{P}}, 
\end{align}
which provide, respectively, the MOG parameter and the mass–to–radius ratio as functions of only directly observable set $\{w_m,h_m,\delta_m\}$ measured on the midline.

It is important to note that we have derived an exact closed-form expression for the MOG coupling $\alpha$ solely in terms of directly observable quantities, without introducing redshift rapidity. Substituting $\tilde{P}$ and $\tilde{N}$ from Eqs. (\ref{tildeM})-(\ref{tildeQ}) in relation (\ref{m}), we obtain the following explicit expression for $\alpha$
\begin{widetext}
\begin{eqnarray}
\label{alphade}   
\alpha =
\frac{\dfrac{8}{w_m}-\dfrac{12+3w_m\tan^2{\delta_m}}{4h_m}+3\sec^2{\delta_m}-2}
{\left(\dfrac{4}{w_m}-\dfrac{4+w_m\tan^2{\delta_m}}{2h_m}+2\sec^2{\delta_m}-1\right)^{2}
-\left(\dfrac{8}{w_m}-\dfrac{12+3w_m\tan^2{\delta_m}}{4h_m}+3\sec^2{\delta_m}-2\right)},
\end{eqnarray}
\end{widetext}
where the right-hand side vanishes in the standard Schwarzschild limit $\alpha=0$.\footnote{In the general-relativistic limit $\alpha=0$, one has $\widetilde N=0$ and $\widetilde P=M/r_e$. This can be checked directly at the level of the midline observables. Writing $x=M/r_e$, $S_m=\sec^2\delta_m$, and $T_m=\tan^2\delta_m=S_m-1$, Eqs.~\eqref{rb1}-\eqref{rb2}  reduce to $w_m^{(0)}=4/(1-3x)$ and $h_m^{(0)}=(S_m-3x)/[(1-3x)(S_m-2x)]$. Substituting these expressions into the numerator of Eq.~\eqref{alphade} gives
\[
\frac{8}{w_m^{(0)}}
-
\frac{12+3w_m^{(0)}T_m}{4h_m^{(0)}}
+3S_m-2
=
2(1-3x)-3(S_m-2x)+3S_m-2=0,
\]
whereas the denominator reduces to $x^2$. Therefore, the right-hand side of Eq.~\eqref{alphade} vanishes, as required.}

In the next section, we introduce the redshift rapidity and disentangle $M/r_e$ in the Schwarzschild-MOG background.

\subsection{Redshift at the line of sight}

The second important case is describing the frequency shift of photons emitted close to the LOS where $\varphi \rightarrow 0$. Hence, by substituting $\varphi =\varphi
_{s}\sim 0$\ in the frequency shift formula (\ref{Redshift}), we find the expressions for slightly redshifted and slightly blueshifted photons as below%
\begin{equation}\label{losfreq}
1+z_{_{MOG_{1,2}}}^{(s)}\!=\sqrt{\frac{1}{1-3\tilde{P} +2\tilde{N}}}\pm \sqrt{\frac{\tilde{P} - \tilde{N}}{%
1-3\tilde{P}+ 2\tilde{N}  }}\left( \varphi _{s}+\delta _{s}\right) ,
\end{equation}%
where the index \textquotedblleft $s$"\ means the measurement\ should be
performed for systemic particles (particles close to the LOS) and
we applied the limit $\delta _{s}\rightarrow 0$\ simultaneously. Additionally, we
have just the gravitational redshift $z_{g}$ exactly at the LOS
with $\varphi _{s}=0=\delta _{s}$, which is an increasing function of the MOG parameter $\alpha$ (see Lemma~\ref{ZgLem} and the left panels of Fig.~\ref{zDotFig}). In this relation, the angles $\delta _{s}$%
\ and $\varphi _{s}$\ should be measured close to the LOS, and the
plus (minus) sign refers to the redshifted (blueshifted) photons close to the LOS denoted by $z_{_{MOG_{1}}}^{(s)}$ ($z_{_{MOG_{2}}}^{(s)}$).
Now, defining $w_{s}:=(2+z_{_{MOG_{1}}}^{(s)}   +z_{_{MOG_{2}}}^{(s)})^2$ from (\ref{losfreq}), we obtain [analogous to Eq. (\ref{rb2})]

\begin{equation}
w_s= \frac{4}{1-3\tilde{P}+2\tilde{N}}.  \label{EQws}
\end{equation}%

By considering Eq. (\ref{EQws}), one can show that the mass-to-radius ratio close to the line of sight is given by
\begin{equation}
\frac{M\varphi_s}{D \delta_s}=\frac{3 (\tilde{P}^2-\tilde{N})}{4 \tilde{N}} \left[1- \sqrt{1-\frac{8 \tilde{N}}{9 \tilde{P}^2}\left(1-\frac{4}{w_s}  \right)} \right],
\label{MtoDLOS}
\end{equation}
in terms of directly measurable quantities. To derive this relation, we have used $r_e \varphi_s=D \delta_s$ valid close to the line of sight (see relation (\ref{rel2})) and substituted $\alpha$ from (\ref{m}). Note that in Eq. (\ref{MtoDLOS}), $\tilde{P}$ and $\tilde{N}$ are given in (\ref{tildeM})-(\ref{tildeQ}) and they should be measured on the midline, whereas the set \{$\varphi_s$, $\delta_s$, $w_s$\} should be measured close to the LOS.

\section{REDSHIFT RAPIDITY IN THE SCHWARZSCHILD--MOG BACKGROUND}\label{sec:rapidity}

In this section, we extend the formalism to write the black hole mass \(M\) and its distance \(D\) from the observer purely in terms of directly observable quantities (the parameter \(\alpha\) has already been isolated). To accomplish this we use the \emph{redshift rapidity}, a notion recently introduced for the Schwarzschild spacetime \cite{Momennia2024EPJC}.

We define the redshift rapidity as the proper–time rate of change of the frequency shift \(z_{MOG}\) in (\ref{Redshift}) in the MOG background,
\begin{equation}\label{redrapi}
      \dot{z}_{MOG,e}=\frac{dz_{MOG}}{d\tau}=\frac{d}{d\tau}\!\left(U^t_e-b_\gamma U^\varphi_e\right),
\end{equation}
evaluated at the emission event. Since the quantity of interest must be determined at the observer's location, we apply the chain rule to recast (\ref{redrapi}) at the detection point \cite{Momennia2024EPJC}:
\begin{equation}\label{redrapit}
    \dot{z}_{MOG}=\frac{dz_{MOG}}{dt}=\frac{d\tau}{dt}\frac{dz_{MOG}}{d\tau}
    =\frac{1}{U^t_e}\frac{d}{d\tau}\!\left(U^t_e-b_\gamma U^\varphi_e\right),
\end{equation}
which is an observational quantity measured on Earth, and we have used $U^t_d=1$ in the first step according to Eq. (\ref{Detector4Velocity}). For test particles on circular geodesics in the equatorial plane, this becomes
\begin{equation}
    \dot{z}_{MOG}=-\frac{db_\gamma}{d\tau}\,\frac{U^\varphi_e}{U^t_e},
\end{equation}
because \(U^t_e\) in (\ref{Ut}) and \(U^\varphi_e\) in (\ref{Uphi}) are constants for circular motion, while the impact parameter (\ref{lbp}) varies through its dependence on \(\delta\) and \(\varphi\). Employing the chain rule, one finds
\begin{equation}\label{totredrapi}
    \dot{z}_{MOG}=-\left(\frac{\partial b_\gamma}{\partial\varphi}
    +\frac{\partial b_\gamma}{\partial\delta}\frac{\partial\delta}{\partial\varphi}\right)\frac{(U^\varphi_e)^2}{U^t_e},
\end{equation}
where we used \(U^\varphi_e=\frac{d\varphi}{d\tau}\big|_{r=r_e}\). Carrying out the derivatives of (\ref{lbp}) and of \(\delta(\varphi)\) in (\ref{deltaphirel}), the redshift rapidity at a generic point on the orbit is
\begin{align}\label{RRgeneral}
    \dot{z}_{MOG}=&\frac{D}{r_e}\left[\frac{1-\tilde{P}-g(r_e)}{\sqrt{2g(r_e)+\tilde{P}-1}}\right]
    \left[\frac{D-r_e\cos{\varphi}}{r_e^2+D^2-2r_eD\cos{\varphi}}\right]\times\notag \\
    &\frac{\cos(\varphi + \delta)}
    {\left[g(r_e)\sin^2(\varphi+\delta)+\cos^2(\varphi+\delta)\right]^{3/2}},
\end{align}
which correctly reduces to the Schwarzschild expression \cite{Momennia2024EPJC} when \(\alpha=0\).

The right panels of Fig.~\ref{zDotFig} display \(\dot{z}_{MOG}\) as a function of the azimuthal angle \(\varphi\) for several values of the MOG parameter \(\alpha\). Unlike the case of the total frequency shift, the rapidity either increases or decreases with $\alpha$, depending on the emitter radius. In the upper-right panel we set \(r_e=10\,r_{ISCO}(\alpha)\) and \(D=10^{5}r_{ISCO}(\alpha)\); these plots show how the redshift rapidity evolves with the particle's motion. Moreover, the rapidity attains its maximum along the line of sight (\(\varphi=0\)), which facilitates its measurement. It is worth noting that these figures illustrate the projection of the redshift rapidity on the LOS, just like the frequency shift case.

Likewise, in the weak-field (Newtonian) limit in which $M/r_e \to 0$ ,
the \emph{redshift rapidity} (\ref{RRgeneral}) at leading order simplifies to the line-of-sight projection
of the Keplerian acceleration for a massive test particle on a circular orbit:
\begin{equation}
\dot z_{\text{Newton}}=(1+\alpha) \frac{M}{r_e^{2}}\cos(\varphi+\delta),
\end{equation}
where $\alpha$ shows the correction due to modified gravity. For a distant observer and in the absence of modified gravity correction, $\alpha=0$, this is precisely the classical result one expects.

\subsection{Redshift rapidity at the midline}

Here, we disentangle the mass–to–radius ratio (\ref{m}) and provide analytic formulas for the black hole mass $M$, distance to the black hole $D$, and the emitter radius $r_e$ in terms of directly measurable quantities. To this end, we substitute \(r_e\) from Eq. (\ref{rel2}) and we set \(\varphi=\varphi_m=\pm\pi/2\) to evaluate the rapidity (\ref{RRgeneral}) on the midline. Considering the absolute value of $\dot{z}_{MOG}^{(m)}$ and solving the resultant relation for $D$ in terms of observables, we obtain
\begin{align}\label{D-decop}
D=\frac{w_m-4 h_m}{\dot{z}_{MOG}^{(m)} \sqrt{w_m \tan^2 \delta_m + 4}}\sqrt{\frac{h_m}{w_m}},
\end{align}
where we have replaced $\tilde{P}$ and $\tilde{N}$ given by Eqs. (\ref{tildeM})-(\ref{tildeQ}). Now, inserting $r_e=D \tan \delta_m$ and Eq. (\ref{D-decop}) into the mass–to–radius ratio (\ref{m}) yields the total mass of the Schwarzschild--MOG black hole:
\begin{widetext}
\begin{eqnarray}
M&=&\frac{w_m-4 h_m}{\dot{z}_{MOG}^{(m)} \sqrt{4 \cot^2 \delta_m +w_m}}\sqrt{\frac{h_m}{w_m}} \times \notag\\
&&\left(2 \sec ^2 \delta_m-\frac{w_m \tan ^2 \delta_m +4}{2 h_m}+\frac{4}{w_m}-1-\dfrac{3 \sec ^2 \delta_m -\dfrac{3 w_m \tan ^2 \delta_m +12}{4 h_m}+\dfrac{8}{w_m}-2}{2 \sec^2 \delta_m -\dfrac{w_m \tan^2 \delta_m +4}{2 h_m}+\dfrac{4}{w_m}-1}\right).  
\label{M-decop}
\end{eqnarray}
\end{widetext}

Besides, one can express the emitter radius $r_e$ in terms of observables as follows
\begin{equation}
r_e=\frac{w_m-4 h_m}{\dot{z}_{MOG}^{(m)} \sqrt{4 \cot^2 \delta_m +w_m}} \sqrt{\frac{h_m}{w_m}}.
\label{reDecop}
\end{equation}

Recall that \(\alpha\) has already been expressed in terms of observables in Sec.~\ref{sec:freqshift}. Therefore, from (\ref{alphade}), (\ref{D-decop}), (\ref{M-decop}), and (\ref{reDecop}), it follows that the MOG parameter \(\alpha\), distance \(D\) to the black hole, total mass \(M\), and the radius of emitter $r_e$ are determined solely by the set of directly measured quantities \{$h_m$, $w_m$, $\dot{z}_{MOG}^{(m)}$, $\delta_m$\} on the midline— namely, the total redshift \(z_{MOG_1}^{(m)}\), total blueshift \(z_{MOG_2}^{(m)}\), redshift rapidity \(\dot{z}_{MOG}^{(m)}\), and the aperture angle \(\delta_m\).

\subsection{Redshift rapidity at the line of sight}

The second important case is related to the test particles that lie close to the LOS where their electromagnetic radiation is slightly
frequency shifted and their rapidity is maximal. The
mass-to-distance ratio of the black hole is derived in Eq. (\ref{MtoDLOS}) in terms of the redshifted/blueshifted photons.

On the other hand, close to the LOS, where $\varphi=\varphi_s \approx 0$ and $\delta=\delta_s \approx 0$, the redshift rapidity (\ref{RRgeneral}) reduces to
\begin{equation}
\dot{z}_{MOG}^{(s)}=\frac{D (\tilde{P}-\tilde{N})}{r_e (D-r_e) \sqrt{2 \tilde{N}-3 \tilde{P}+1}}+\mathcal{O}(\varphi_s \delta_s),
\end{equation}
where we discarded the quadratic and higher-order terms in $\varphi_s$ and $\delta_s$. Then, we substitute $r_e=MD \delta_s /(M \varphi_s)$ from Eq. (\ref{rel2}) to obtain
\begin{equation}
\dot{z}_{MOG}^{(s)}=\frac{D (\tilde{P}-\tilde{N})}{\frac{D \delta_s}{M \varphi_s} M \left( D-\frac{D \delta_s}{M \varphi_s} M \right) \sqrt{2 \tilde{N}-3 \tilde{P}+1}}. 
\end{equation}

Finally, we insert $M= (\tilde{P}-\tilde{N}/\tilde{P}) D \tan \delta_m$ and $\frac{M \varphi_s}{D \delta_s}$, respectively, from the relations (\ref{m}) and (\ref{MtoDLOS}), and solve for $D$ to get\footnote{Note that the emitter radius $r_e \approx D \tan \delta_m \approx D \delta_s/\varphi_s$ is a constant for a particular orbit.}
\begin{eqnarray}
D&=&\frac{3 \tilde{P} (\tilde{P}-\tilde{N})\cot \delta_m}{2 \tilde{N} \dot{z}_{MOG}^{(s)} \sqrt{1-3 \tilde{P}+2 \tilde{N}}} \times \notag \\
&&\left[\frac{1-\sqrt{1+\frac{8 \tilde{N}}{9 \tilde{P}^2} \left(\frac{4}{w_s}-1\right)}+\frac{4 \tilde{N}}{9 \tilde{P}^2}\left(\frac{4}{w_s}-1\right)}{1-\sqrt{1+\frac{8 \tilde{N}}{9 \tilde{P}^2} \left(\frac{4}{w_s}-1\right)}-\frac{4 \tilde{N}}{3 \tilde{P} \cot \delta_m}}\right],\qquad \label{Dlos}
\end{eqnarray}
for the distance to the black hole in terms of directly measurable quantities on the midline $\{w_m,h_m,\delta_m\}$ and close to the line-of-sight $\left\{w_s,\dot{z}_{MOG}^{(s)},\delta_s\right\}$. From an observational point of view, Eq. (\ref{Dlos}) is an interesting relation for measuring the distance to a Schwarzschild-MOG black hole because it consists of a combination of highly redshifted particles on the midline and maximal rapidity at the LOS, where these observables are easier to be measured.

As the final remark regarding the static Schwarzschild--MOG black hole, we would like to stress that the analytic formulas (\ref{alphade}), (\ref{D-decop}), (\ref{M-decop}), (\ref{reDecop}), and (\ref{Dlos}) are among the main results of this article. These exact, analytic relations are valid on the midline and close to LOS, and they can be directly applied to black hole systems located at the center of galaxies. On the other hand, the exact formulas of the redshift~(\ref{Redshift}) and redshift rapidity (\ref{RRgeneral}) are valid on the whole circular orbit, and they are important in black hole parameter estimation studies.

\section{Kerr--MOG extension}\label{sec:kerrmog}

In this section, we further extend previous results from a general
relativistic method to obtain the frequency shift, \emph{redshift rapidity}, and \emph{redshift acceleration} formulas of massive probe
particles revolving in Kerr-MOG spacetime background, and then express the black hole parameters of the associated Kerr-MOG black hole in terms of directly measurable quantities. We first
consider the geodesic motion of massive test particles orbiting a Kerr-MOG black
hole which emit photons toward a distant observer, similar to the previous static case. In this section, we work based on the total redshift, which is a directly measurable quantity. Besides, our relations are obtained for an arbitrary point in the circular motion, and we incorporated redshift rapidity, redshift acceleration, and aperture angle of the telescope into our formalism. 

The spacetime background of the rotating black hole is described by the line
element~\cite{Moffat2015EPJC}
\begin{equation}
ds^{2}=g_{tt}dt^{2}+2g_{t\varphi }dtd\varphi +g_{\varphi \varphi }d\varphi
^{2}+g_{rr}dr^{2}+g_{\vartheta \vartheta }d\vartheta ^{2},  \label{metric}
\end{equation}%
with the metric components 
\begin{eqnarray}
g_{tt}&=&-\frac{\Delta-a^2 \sin^2 \vartheta}{\rho^2},\quad g_{rr}=\frac{\rho^2}{\Delta},\\
g_{t\varphi}&=&-\frac{r^2+a^2-\Delta}{\rho^2}a\sin^2\vartheta, \quad g_{\vartheta \vartheta }=\rho^2, \\
g_{\varphi \varphi }&=&\frac{(r^{2}+a^{2})^2-\Delta a^{2}\sin ^{2}\vartheta}{%
\rho^2 } \sin ^{2}\vartheta,
\end{eqnarray}
where $\Delta =r^{2}+a^{2}-2(1+\alpha)Mr + \alpha(1+\alpha)M^2$ and $\rho^2 =r^{2}+a^{2}\cos ^{2}\vartheta$.
In addition, $M$ is the total mass of the Kerr-MOG black hole and $a$ is its total
angular momentum per unit mass, $a=J/M$ ($0\leq a\leq M$). The Kerr-MOG
spacetime has Cauchy horizon $r_-$ and event horizon $r_+$ surfaces located at
\begin{equation}
r_\pm=(1+\alpha)M\biggl[1\pm\sqrt{1-\frac{a^2}{(1+\alpha)^2M^2}-\frac{\alpha}{1+\alpha}}\biggr].
\end{equation}

Besides an ergosphere horizon is determined by $g_{tt}=0$
\begin{equation}
r_E=(1+\alpha)M\biggl[1+\sqrt{1-\frac{a^2\cos^2\vartheta}{(1+\alpha)^2M^2}-\frac{\alpha}{1+\alpha}}\biggr].
\end{equation}

\subsection{Redshift in the Kerr-MOG background}

Performing a similar procedure as the one followed in Section \ref{sec:geodesics}, we obtain the relevant conserved quantities in this Kerr-MOG spacetime as a function of the black hole parameters~\cite{KerrMOG18}
\begin{eqnarray}
E\left(r_e,\pi/2\right) &=& 
\frac{1-2\tilde{P}+\tilde{N}\pm \tilde{a} \sqrt{\tilde{P}-\tilde{N}}}{\left( 1-3\tilde{P}+2\tilde{N}\pm 2\tilde{a} \sqrt{\tilde{P}-\tilde{N}} \right)^{\frac{1}{2}}}, \label{E_KerrMOG} \\
L\left(r_e,\pi/2\right)&=& 
(\pm) \frac{ (1+\tilde{a}^2) \sqrt{\tilde{P}-\tilde{N}} \mp \tilde{a} (2\tilde{P}-\tilde{N}) }{\left( 1-3\tilde{P}+2\tilde{N}\pm 2\tilde{a} \sqrt{\tilde{P}-\tilde{N}} \right)^{\frac{1}{2}}}r_e, \label{L_KerrMOG} \qquad
\end{eqnarray}
where reduce to the corresponding relations (\ref{energy})-(\ref{energy2}) for the vanishing rotation parameter $a/r_e=\tilde{a}=0$. In what follows, we
focus on the clockwise rotation of the emitter only (plus
sign enclosed in parentheses of $L$ relation) and discard the counterclockwise motion without
loss of generality. The upper sign applies to a co-rotating emitter (i.e., the angular momentum of the massive test particle is aligned with that of the Kerr-MOG black hole), whereas the lower sign corresponds to a counter-rotating one.

In axially symmetric spacetimes of the form (\ref{metric}), the frequency
shift of photons emitted by massive geodesic particles orbiting the black
hole and detected by an observer is given by~(\ref{GeneralShift}). Now, because of the existing data regarding the accretion disks as well as
being able to extract analytic formulas for mass, spin, MOG parameter, and distance to the
Kerr-MOG black hole, we restrict ourselves to the circular motion of the photon
sources ($U_{e}^{r}=0$). In addition, we put
both the emitter and observer in the equatorial plane ($\vartheta =\pi /2$),
for which we have $U_{e}^{\vartheta }=0=U_{d}^{\vartheta }$. In the special case of
a distant observer $r_{d}\rightarrow \infty $, the $4$-velocity of the detector
simplifies to $U_{d}^{\mu }=\delta _{t}^{\mu }$ as in the Schwarzhild-MOG case. By taking into account these assumptions, which is the case for the
real astrophysical systems in AGNs, the generic
relation~(\ref{GeneralShift}) reduces to 
\begin{equation}
1+z_{KMOG}\!=\frac{\left. \left( E_{\gamma }U^{t}-L_{\gamma }U^{\varphi
}\right) \right\vert _{e}}{\left. \left( E_{\gamma }U^{t}\right) \right\vert
_{d}}=U_{e}^{t}-b_{\varphi }\,U_{e}^{\varphi }\,,  \label{zcircorbits}
\end{equation}%
where $b_{\varphi }\equiv L_{\gamma }/E_{\gamma }$ is the deflection of
light parameter which gives the light bending produced by the gravitational
field of the Kerr-MOG black hole. Besides, $E_{\gamma }$ and $L_{\gamma }$ are conserved along the
light trajectories and correspond to the total energy and axial angular
momentum of the photons, respectively. The nonvanishing components of the $4$%
-velocity $U_{e}^{\mu }$ are given by~\cite{KerrMOG18} 
\begin{eqnarray}
U^{t}_{e}&=& 
\frac{1\pm \tilde{a} \sqrt{\tilde{P}-\tilde{N}}}
{\left( 1-3\tilde{P}+2\tilde{N}\pm 2\tilde{a} \sqrt{\tilde{P}-\tilde{N}} \right)^{\frac{1}{2}}}, 
\label{eq:Ut_midline_c}\\
U^{\varphi}_{e}&=& \pm
\frac{\sqrt{\tilde{P}-\tilde{N}}}
{r_e \left( 1-3\tilde{P}+2\tilde{N}\pm 2\tilde{a} \sqrt{\tilde{P}-\tilde{N}} \right)^{\frac{1}{2}}}. \quad
\label{eq:Up_midline_c}
\end{eqnarray}

One may also compute the $(\varphi +\delta )$%
-dependent light bending parameter $b_{\varphi }$\ for an arbitrary point of
the circular orbit of photon sources on the equatorial plane, following the procedure of Ref. \cite{Momennia:2025:KerrParams}
\begin{eqnarray}
b_\varphi&=&-\frac{2\tilde{P}-\tilde{N}}{1-2\tilde{P}+\tilde{N}} \tilde{a} r_e
-\frac{r_e \tilde{\Delta}^{3/2} \sin(\varphi+\delta)}{1-2\tilde{P}+\tilde{N}} \times \notag \\
&& \frac{1}{\sqrt{\tilde{\Delta}^2 \sin^2(\varphi+\delta)+(1-2\tilde{P}+\tilde{N})\cos^2(\varphi+\delta)}},\qquad
\label{eq:bphi_c}
\end{eqnarray}
where $\tilde{\Delta}=\Delta(r_e)/r_e^2 =1+\tilde{a}^{2}-2 \tilde{P} + \tilde{N}$ and we used $g_{t \varphi}^2-g_{t t}g_{\varphi \varphi}=\Delta$. Besides, $\varphi $ is the
azimuthal angle that is not a measurable quantity, and $\delta $ is the
aperture angle of the telescope (angular distance) that is an observable
parameter as seen with the Schwarzhild-MOG black hole.

\begin{figure*}[t]
\centering
\includegraphics[width=\textwidth]{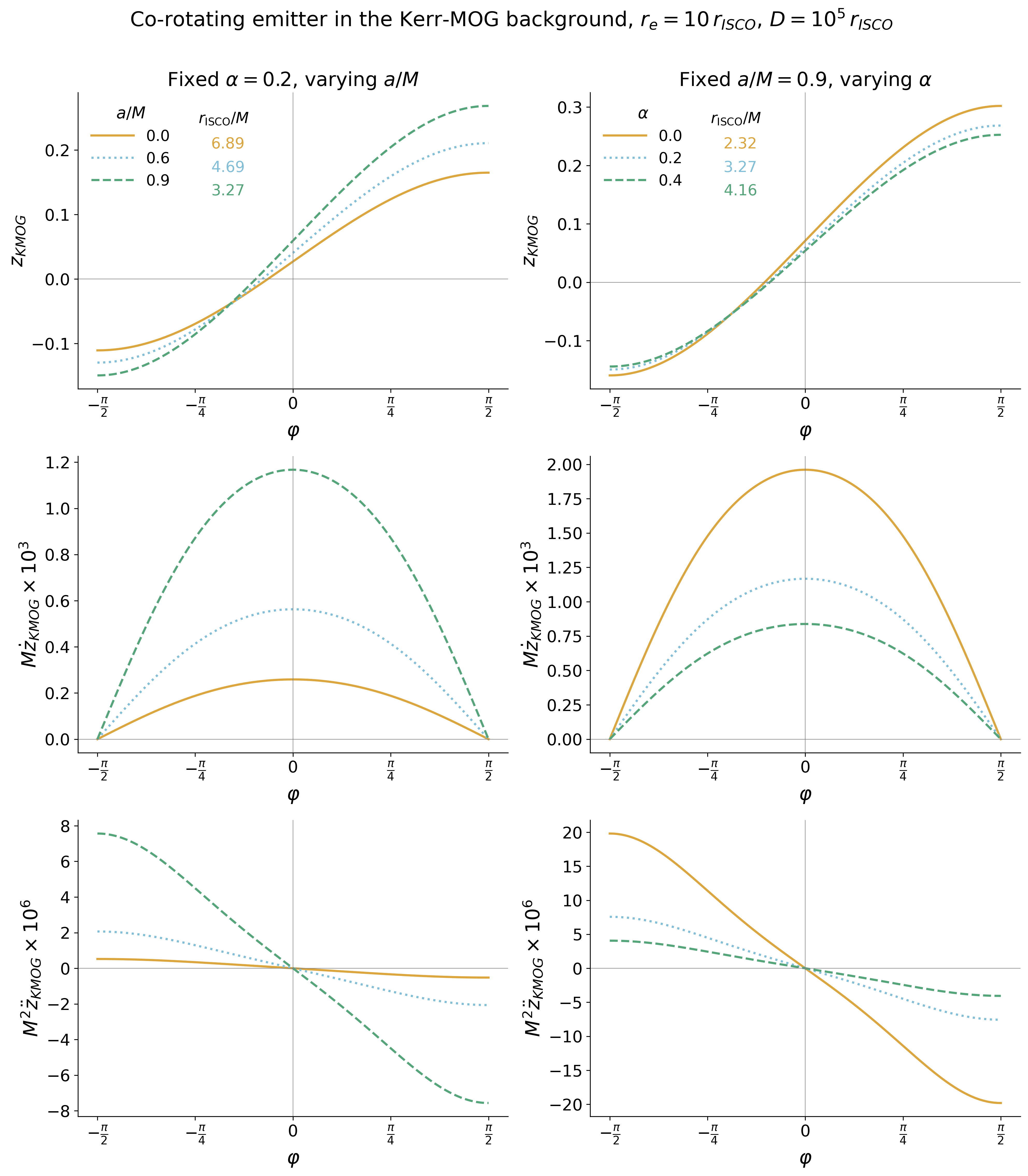}
\caption{
The frequency shift $z_{KMOG}$ (top panels), redshift rapidity $M\dot z_{KMOG}$ (middle panels), and redshift acceleration $M^{2}\ddot z_{KMOG}$ (bottom panels) versus the azimuthal angle $\varphi$ in the Kerr-MOG black hole spacetime. In all panels we take $r_e=10\,r_{\mathrm{ISCO}}$ and $D=10^{5}r_{\mathrm{ISCO}}$, where $r_{\mathrm{ISCO}}$ is the largest real solution of the Kerr-MOG ISCO equation~(\ref{eq:isco_cubic_mog}). In the left column, the MOG parameter is fixed at $\alpha=0.2$, and the curves correspond to different values of the spin parameter $a/M=0.0,\,0.6,$ and $0.9$; in particular, the bottom-left panel shows the redshift acceleration for fixed $\alpha=0.2$ and varying $a/M$. In the right column, the spin is fixed at $a/M=0.9$, and the curves correspond to different values of the MOG parameter $\alpha=0.0,\,0.2,$ and $0.4$; in particular, the bottom-right panel shows the redshift acceleration for fixed $a/M=0.9$ and varying $\alpha$. The plots are shown for the co-rotating branch, corresponding to the $+\tilde a$ choice in the formulas (\ref{zKerrMOG}), (\ref{eq:redshift_rapidity_MOG}), and (\ref{eq:redshift_acceleration_MOG}), and we used (\ref{deltaphirel}) as well. The continuous curves indicate the Schwarzschild-MOG black hole (left panels) and the standard Kerr black hole (right panels).
}
\label{zDotFig2}
\end{figure*}

\clearpage

Now, by substituting (\ref{eq:Ut_midline_c})-(\ref{eq:bphi_c}) into (\ref%
{zcircorbits}), we obtain the following explicit form of the frequency shift
for an arbitrary point of the orbit on the equatorial plane
\begin{widetext}
\begin{eqnarray}
1+z_{KMOG}&=&\frac{1}{\left( 1-3\tilde{P}+2\tilde{N}\pm 2\tilde{a} \sqrt{\tilde{P}-\tilde{N}} \right)^{\frac{1}{2}}} \times \notag \\
&&\left[ 1 \pm \frac{
\sqrt{\tilde{P}-\tilde{N}}}{1-2\tilde{P}+\tilde{N} }\left(\tilde{a} +   \frac{\tilde{\Delta}^{3/2} \sin(\varphi+\delta)}{\sqrt{\tilde{\Delta}^2 \sin^2(\varphi+\delta)+(1-2\tilde{P}+\tilde{N})\cos^2(\varphi+\delta)}}\right) \right],
\label{zKerrMOG}
\end{eqnarray}
\end{widetext}
where we recall $\tilde{P}=(1+\alpha)M/r_{e}$, $\tilde{N}=\alpha(1+\alpha)M^2/r_{e}^2$, and $\ \tilde{a}=a/r_{e}$. Within this relation, the upper (lower) sign corresponds to co-rotating (counter-rotating) photon sources. One notes that this relation reduces to the frequency shift in the standard Kerr spacetime in the limit $\alpha=0$~\cite{Momennia:2025:KerrParams}.

Besides, the ISCO radius in the Kerr-MOG
spacetime is the largest real value solution of the cubic equation \begin{align}
0 ={}&
M r_{\rm ISCO}^3
- 6(1+\alpha) M^2 r_{\rm ISCO}^2
- 3M a^2 r_{\rm ISCO}
\notag\\[2pt]
&{}+ 9\alpha(1+\alpha) M^3 r_{\rm ISCO}
+ 4\alpha M^2
   \big[a^2 - \alpha(1+\alpha)M^2\big]
   \notag\\[2pt]
&{} \mp 8 a \big[(1+\alpha)^\frac{1}{3}(M r_{\rm ISCO}
           - \alpha M^2) \big]^{3/2},
\label{eq:isco_cubic_mog}
\end{align}
that approximately characterizes the inner edge of the accretion disk. In
this article, we are interested in stable circular orbits of the photon
sources such that $r_{e}\geq r_{ISCO}$.

The top panels of Fig.~\ref{zDotFig2} display the frequency shift $z_{\rm KMOG}$ as a function of the azimuthal angle $\varphi$ in the Kerr--MOG background, for the co-rotating branch. As in the Schwarzschild--MOG case, the largest redshift and blueshift occur near the midline, $\varphi \simeq \pm \pi/2$, while the shift changes sign close to the line of sight. In the left panel, with $\alpha$ fixed, increasing the rotation parameter $\tilde a$ enhances the separation between the redshifted and blueshifted sides of the orbit. By contrast, in the right panel, with $\tilde a$ fixed, increasing the MOG parameter $\alpha$ produces the opposite tendency in the profile. Therefore, the effect of the rotation parameter is opposite to the MOG parameter effect on the total frequency shift. In the limits $a \to 0$ and $\alpha \to 0$, one recovers the Schwarzschild--MOG and standard Kerr cases, respectively.

\subsection{Redshift rapidity in the Kerr-MOG background}

In this section, analogous to the Schwarzschild-MOG section, we derive a closed form relationship for the redshift rapidity in a Kerr-MOG background. Substituting Eqs. (\ref{deltaphirel}) and (\ref{eq:Ut_midline_c})-(\ref{eq:bphi_c}) into (\ref{totredrapi}) and simplifying, we obtain
    
\begin{widetext}
\begin{equation}
\label{eq:redshift_rapidity_MOG}
\begin{aligned}      
\dot z_{\mathrm{KMOG}}
&=
\frac{D}{r_e}
\left[
\frac{(\tilde P-\tilde N)\,\tilde\Delta^{3/2}}
{\left(1\pm \tilde a\sqrt{\tilde P-\tilde N}\right)
\sqrt{1-3\tilde P+2\tilde N\pm 2\tilde a\sqrt{\tilde P-\tilde N}}}
\right]
\times
\\[4pt]
&\qquad
\left[
\frac{D-r_e\cos\varphi}{D^2+r_e^2-2Dr_e\cos\varphi}
\right]
\frac{\cos(\varphi+\delta)}
{\left[
\tilde\Delta^2\sin^2(\varphi+\delta)
+
\left(1-2\tilde P+\tilde N\right)\cos^2(\varphi+\delta)
\right]^{3/2}},
\end{aligned}
\end{equation}
\end{widetext}
where reduces to the frequency shift in the standard Kerr spacetime in the limit $\alpha=0$~\cite{Momennia:2025:KerrParams}.

The middle panels of Fig.~\ref{zDotFig2} display the redshift rapidity $M\dot z_{\rm KMOG}$ as a function of the azimuthal angle $\varphi$ in the Kerr--MOG background. As in the Schwarzschild--MOG case, the rapidity attains its maximum along the line of sight, $\varphi = 0$, which facilitates its measurement. In the left panel, for fixed $\alpha$, increasing the spin parameter $a/M$ increases the amplitude of the rapidity. On the other hand, for fixed $a/M$ in the right panel, increasing the MOG parameter $\alpha$, decreases it. Hence, the effect of the rotation parameter is opposite to the MOG parameter effect also for the redshift rapidity. As in the non-rotating case, these figures illustrate the projection of the redshift rapidity on the LOS, now in the Kerr--MOG background.

\subsection{Redshift acceleration in the Kerr-MOG background}

Following the derivation of the redshift rapidity, we now calculate the \textit{redshift acceleration}, defined as the proper time derivative of the redshift rapidity $\dot{z}_{KMOG}$. This quantity provides a deeper probe into the higher-order kinematic effects in the Kerr-MOG background.

Starting from the redshift rapidity expression in Eq. (\ref{totredrapi}), we differentiate with respect to the coordinate time $t$ in order to obtain the redshift acceleration. Using the relation $dt = U^t_e d\tau$, we can express this in terms of the proper time $\tau$:

\begin{equation}
\ddot{z}_{KMOG} = \frac{d\dot{z}_{KMOG}}{dt} =  \frac{1}{U^t_e}\frac{d}{d\tau} \left[ \dot{z}_{KMOG} \right],
\label{redAccFirst}
\end{equation}
evaluated at the emission event. Substituting Eq. (\ref{totredrapi}) into this expression, and recalling that for circular geodesics the four-velocity components $U^t_e$ and $U^\varphi_e$ are constants, the derivative acts only on the impact parameter term. Introducing Eq. (\ref{totredrapi}) into (\ref{redAccFirst}), we obtain:
\begin{equation}
\ddot{z}_{KMOG} = - \frac{(U^\varphi_e)^2}{(U^t_e)^2} \frac{d}{d\tau} \left( \frac{\partial b_\gamma}{\partial \varphi} + \frac{\partial b_\gamma}{\partial \delta}\frac{\partial \delta}{\partial \varphi} \right).
\end{equation}

Carrying out the derivative with respect to $\tau$, using the chain rule and expanding, we arrive at the final expression for the redshift acceleration measured by a distant observer:
\begin{equation}
\begin{aligned}
\ddot{z}_{KMOG} = & - \frac{(U^\varphi_e)^3}{(U^t_e)^2} \Bigg[ \frac{\partial^2 b_\gamma}{\partial \varphi^2} + 2\frac{\partial^2 b_\gamma}{\partial \varphi \partial \delta}\frac{\partial \delta}{\partial \varphi} \\
& + \frac{\partial^2 b_\gamma}{\partial \delta^2}\left(\frac{\partial \delta}{\partial \varphi}\right)^2 + \frac{\partial b_\gamma}{\partial \delta}\frac{\partial^2 \delta}{\partial \varphi^2} \Bigg],
\label{RedAccThird}
\end{aligned}
\end{equation}
where we have assumed the symmetry of mixed partial derivatives and grouped the terms accordingly. Before employing this result to decouple the system $ \{M,a,\alpha,D\}$, it is useful to obtain the closed formula for the redshift acceleration in terms of black hole parameters with the help of Eqs. (\ref{deltaphirel}) and (\ref{eq:Ut_midline_c})-(\ref{eq:bphi_c}), and (\ref{RedAccThird}):

\begin{widetext}
\begin{equation}
\label{eq:redshift_acceleration_MOG}
\begin{aligned}
\ddot z_{KMOG}
&=
\mp
\frac{(\tilde P-\tilde N)^{3/2}\,\tilde\Delta^{3/2}}
{\left(1\pm \tilde a\sqrt{\tilde P-\tilde N}\right)^2
\sqrt{1-3\tilde P+2\tilde N\pm 2\tilde a\sqrt{\tilde P-\tilde N}}}
\times
\\[4pt]
&\quad
\Bigg\{
\frac{D(D^2-r_e^2)\sin\varphi\,\cos(\varphi+\delta)}
{r_e\left(D^2+r_e^2-2Dr_e\cos\varphi\right)^2
\left[
\tilde\Delta^2\sin^2(\varphi+\delta)
+
\left(1-2\tilde P+\tilde N\right)\cos^2(\varphi+\delta)
\right]^{3/2}}
\\[6pt]
&\qquad\qquad
+
\frac{D^2\left(D-r_e\cos\varphi\right)^2\sin(\varphi+\delta)\,
\Big[
\tilde\Delta^2\sin^2(\varphi+\delta)
+
\left(3\tilde\Delta^2-2\left(1-2\tilde P+\tilde N\right)\right)\cos^2(\varphi+\delta)
\Big]}
{r_e^2\left(D^2+r_e^2-2Dr_e\cos\varphi\right)^2
\left[
\tilde\Delta^2\sin^2(\varphi+\delta)
+
\left(1-2\tilde P+\tilde N\right)\cos^2(\varphi+\delta)
\right]^{5/2}}
\Bigg\}.
\end{aligned}
\end{equation}
\end{widetext}

Likewise, in the weak-field limit, in which $M/r_e \to 0$, $a/r_e \to 0$, and
$D \to \infty$, Eq.~(\ref{eq:redshift_acceleration_MOG})
reduces at leading order to
\begin{equation}
\ddot z_{\mathrm{Newton}}
=
-\Omega^{3}r_{e}\sin(\varphi+\delta)
=
-\frac{\big[(1+\alpha)M\big]^{3/2}}{r_{e}^{7/2}}\sin(\varphi+\delta),
\end{equation}
where \(\Omega=\sqrt{(1+\alpha)M/r_{e}^{3}}\) is the Keplerian angular velocity in the MOG-modified Newtonian regime. Thus, at leading order, the redshift acceleration is the line-of-sight projection of the Newtonian jerk for a particle in circular motion. The MOG correction enters through the replacement \(M\to(1+\alpha)M\), while the dependence on the spin parameter \(a\) appears only beyond leading order term. In the limit \(\alpha\to0\), one recovers the standard Keplerian result. 

This interpretation for the \emph{redshift acceleration} is consistent with the relativistic kinematical hierarchy, in which jerk is the derivative of the acceleration \cite{Russo2009JPA}. It is also of observational interest: in exoplanet time-series analyses, the jerk term becomes relevant once the time baseline probes a sufficiently large fraction of the orbit \cite{Ford2011Bayesian}; in pulsar timing, Dutta \textit{et al.} reported a large and increasing jerk in NGC~1851A and interpreted it as evidence for an ongoing three-body encounter \cite{Dutta2025AA}; and, in cosmology, derivatives of redshift observables can be used to constrain the jerk parameter and thereby test \(\Lambda\)CDM against alternative cosmological models \cite{Martins2016PRD}.

Even though the relativistic jerk is not a directly observational quantity for the system under study, it could be inferred in principle numerically from a set of measurements of the redshift rapidity at different positions in the circular orbit. For this reason, in what follows, we refer to the redshift acceleration as a measurable quantity.

The bottom panels of Fig.~\ref{zDotFig2} display the redshift acceleration $M^2\ddot z_{\rm KMOG}$ as a function of the azimuthal angle $\varphi$ in the Kerr--MOG background. In the left panel, for fixed $\alpha$, increasing the rotation parameter $\tilde a$ significantly enhances the magnitude of $M^2\ddot z_{\rm KMOG}$. By contrast, in the right panel, for fixed $\tilde a$, increasing the MOG parameter $\alpha$ reduces its magnitude. Therefore, the effect of the rotation parameter is opposite to the MOG parameter effect also on the redshift acceleration.

\subsection{Decoupling of parameters}
\label{DecoupParam}

For the remainder of the discussion we limit our attention to co-rotating emitters, as the analysis for counter-rotating ones is straightforwardly obtained by applying the same sequence of steps. Analyzing the system near the midline  where \(\varphi=\varphi_m=\pm\pi/2\) and \(\delta=\delta_m\) (with the redshift  acceleration evaluated at $+\pi/2 $ for simplicity), we obtain a $5\times5$ nonlinear system of equations from the relations (\ref{rel2}), (\ref{zKerrMOG}), (\ref{eq:redshift_rapidity_MOG}), and (\ref{eq:redshift_acceleration_MOG}):

\begin{equation}
r_{e} = D \tan\delta_{m},
\label{eq:re}
\end{equation}

\begin{widetext}

\begin{equation}
\begin{aligned}
1 + z_{KMOG_{1}}^{(m)} =&
\frac{1}{\sqrt{
1 - 3\tilde{P} + 2\tilde{N}
+ 2\tilde{a}\sqrt{\tilde{P} - \tilde{N}}
}}
\\[0.5em]
&\times
\left[
1 + \frac{\sqrt{\tilde{P} - \tilde{N}}}{1 - 2\tilde{P} + \tilde{N}}
\left(
\tilde{a}
+
\frac{\tilde{\Delta}^{3/2}\cos\delta_{m}}
{\sqrt{
\tilde{\Delta}^{2}\cos^{2}\delta_{m}
+
(1 - 2\tilde{P} + \tilde{N})\sin^{2}\delta_{m}
}}
\right)
\right],
\end{aligned}
\label{eq:zplus}
\end{equation}

\begin{equation}
\begin{aligned}
1 + z_{KMOG_{2}}^{(m)} =&
\frac{1}{\sqrt{ 
1 - 3\tilde{P} + 2\tilde{N}
+ 2\tilde{a}\sqrt{\tilde{P} - \tilde{N}}
}}
\\[0.5em]
&\times
\left[
1 + \frac{\sqrt{\tilde{P} - \tilde{N}}}{1 - 2\tilde{P} + \tilde{N}}
\left(
\tilde{a}
-
\frac{\tilde{\Delta}^{3/2}\cos\delta_{m}}
{\sqrt{
\tilde{\Delta}^{2}\cos^{2}\delta_{m}
+
(1 - 2\tilde{P} + \tilde{N})\sin^{2}\delta_{m}
}}
\right)
\right],
\end{aligned}
\label{eq:zminus}
\end{equation}

\begin{equation}
\begin{aligned}
\dot{z}_{KMOG}^{(m)} =&
\frac{D}{r_{e}}
\left[
\frac{(\tilde{P} - \tilde{N})\tilde{\Delta}^{3/2}}
{
\left(1 + \tilde{a}\sqrt{\tilde{P} - \tilde{N}}\right)
\sqrt{
1 - 3\tilde{P} + 2\tilde{N}
+ 2\tilde{a}\sqrt{\tilde{P} - \tilde{N}}
}
}
\right]
\\[0.5em]
&\times
\left[
\frac{D}{D^{2} + r_{e}^{2}}
\right]
\left[
\frac{\sin\delta_{m}}
{
\left(
\tilde{\Delta}^{2}\cos^{2}\delta_{m}
+
(1 - 2\tilde{P} + \tilde{N})\sin^{2}\delta_{m}
\right)^{3/2}
}
\right],
\end{aligned}
\label{eq:zdotr}
\end{equation}

\begin{equation}
\begin{aligned}
\ddot{z}_{KMOG}^{(m)} ={}&
\left[
\frac{(\tilde{P} - \tilde{N})^{3/2}\tilde{\Delta}^{3/2}}
{
\left(1 + \tilde{a}\sqrt{\tilde{P} - \tilde{N}}\right)^{2}
\sqrt{
1 - 3\tilde{P} + 2\tilde{N}
+ 2\tilde{a}\sqrt{\tilde{P} - \tilde{N}}
}
}
\right]
\\[0.75em]
&\times
\Bigg\{
\frac{D(D^{2} - r_{e}^{2})\sin\delta_{m}}
{
r_{e}(D^{2} + r_{e}^{2})^{2}
\left(
\tilde{\Delta}^{2}\cos^{2}\delta_{m}
+
(1 - 2\tilde{P} + \tilde{N})\sin^{2}\delta_{m}
\right)^{3/2}
}
\\[0.75em]
&\qquad
-
\frac{
D^{4}\cos\delta_{m}
\left[
\tilde{\Delta}^{2}\cos^{2}\delta_{m}
+
\left(3\tilde{\Delta}^{2} - 2 + 4\tilde{P} - 2\tilde{N}\right)\sin^{2}\delta_{m}
\right]
}{
r_{e}^{2}(D^{2} + r_{e}^{2})^{2}
\left(
\tilde{\Delta}^{2}\cos^{2}\delta_{m}
+
(1 - 2\tilde{P} + \tilde{N})\sin^{2}\delta_{m}
\right)^{5/2}
}
\Bigg\},
\end{aligned}
\label{eq:zddotr}
\end{equation}
where one should solve to obtain the set of five unobservable elements $\{M,a,\alpha,r_e,D\}$. In order to solve the system, we found that it is convenient to introduce the following shorthand notations in terms of directly measurable quantities 

\begin{equation}
p_m:=1+\frac{z_{KMOG_{1}}^{(m)}+z_{KMOG_{2}}^{(m)}}{2},
\qquad
d_m:=\frac{z_{KMOG_{1}}^{(m)}-z_{KMOG_{2}}^{(m)}}{2},
\qquad
\dot z_m:=\bigl|\dot z_{\mathrm{KMOG}}^{(m)}\bigr|,
\qquad
\eta_m:=-\frac{\ddot z_{\mathrm{KMOG}}^{(m)}}{\dot z_m^{\,2}} ,
\end{equation}

\begin{equation}
s_m:=\sin\delta_m,
\qquad
c_m:=\cos\delta_m .
\end{equation}

\begin{equation}
H_m:=3+d_m\eta_m,
\qquad
K_m:=\tan^2\delta_m\bigl(c_m^2H_m-1\bigr),
\end{equation}

\begin{equation}
\Phi_m:=\frac{1}{\sqrt{K_mp_m^2-d_m^2s_m^2H_m}},
\qquad
R_m:=s_m\sqrt{\frac{H_m(1-\Phi_m)}{K_m}} .
\end{equation}

\begin{equation}
\mathcal A_m:=1-2\tilde P+\tilde N
=\frac{K_m}{K_mp_m^2-d_m^2s_m^2H_m},
\qquad
\tilde\Delta_m:=1+\tilde a^2-2\tilde P+\tilde N
=\frac{K_m}{\sqrt{K_mp_m^2-d_m^2s_m^2H_m}} .
\end{equation}

\begin{equation}
E_m=\frac{1}{p_m-d_mR_m},
\qquad
\tilde a_m=\sqrt{K_m\Phi_m(1-\Phi_m)},
\qquad
q_m^2:=\tilde P-\tilde N
=\frac{d_m^2s_m^2H_m\Phi_m^3}{(p_m-d_mR_m)^2}.
\end{equation}

\begin{equation}
\tilde P_m=1-\mathcal A_m-q_m^2,
\qquad
\tilde N_m=1-\mathcal A_m-2q_m^2 .
\end{equation}

\end{widetext}

We obtain the decoupled system in terms of observables near the midline as follows [see Appendix~\ref{appendix} for the derivation and more explicit expressions in~(\ref{Dapp})-(\ref{aapp})]:

\begin{equation}
D
=\frac{d_m^2\Phi_m^{3/2}}
{\dot z_m\,s_m\sqrt{H_m}\Bigl[p_m+d_mR_m\bigl(\mathcal A_m-1\bigr)\Bigr]},
\end{equation}

\begin{equation}
r_e=D\tan\delta_m,
\qquad
\alpha=\frac{\tilde N_m}{\tilde P_m^{\,2}-\tilde N_m},
\label{KerrMOGparameter}
\end{equation}

\begin{equation}
M=r_e\left(\tilde P_m-\frac{\tilde N_m}{\tilde P_m}\right),
\qquad
a=\tilde a_m\,r_e 
,
\end{equation}
where the right-hand side of Eq.~\eqref{KerrMOGparameter} for $\alpha$ vanishes in the standard Kerr limit $\alpha=0$.\footnote{In the standard Kerr limit $\alpha=0$, the MOG charge combination vanishes, $\widetilde N=0$. Since $q_{m,(0)}^2=\widetilde P-\widetilde N$ and $\mathcal A_m^{(0)}=1-2\widetilde P+
\widetilde N$, one obtains $q_{m,(0)}^2=\widetilde P$ and $\mathcal A_m^{(0)}=1-2\widetilde P$. Thus the reconstructed numerator in Eq.~\eqref{KerrMOGparameter} becomes
\[
\widetilde N_m^{(0)}=1-\mathcal A_m^{(0)}-2q_{m,(0)}^2
=1-(1-2\widetilde P)-2\widetilde P=0,
\]
while $\widetilde P_m^{(0)}=1-\mathcal A_m^{(0)}-q_{m,(0)}^2=\widetilde P$. Hence, the right-hand side of Eq.~\eqref{KerrMOGparameter} also vanishes in the general-relativistic limit.}

These exact analytic formulas are among the main results of the present study and they represent the mass and rotation parameters of the Kerr-MOG black hole, its distance from the Earth, MOG parameter, and the radius of the photon source in terms of directly observational quantities at the most important location of the orbit on the midline. These equations have direct application to supermassive black holes hosted at the center of active galaxies orbited by massive test particles~\cite{MCPII,MCPXI}.  On the other hand, the generic relations (\ref{rel2}), (\ref{zKerrMOG}), (\ref{eq:redshift_rapidity_MOG}), and (\ref{eq:redshift_acceleration_MOG}) find astrophysical applications for modeling the accretion disks revolving supermassive black
holes (see~\cite{ApJL,TXS,TenAGNs, FiveAGNs,SdS} for the standard Schwarzschild and Schwarzschild--de Sitter black hole modelings). It is important to note that we have found a closed form expression for the $\alpha$ parameter in terms of directly observable physical quantities of the accretion disk in a Kerr-MOG background which allows us to directly measure deviations from general relativity.

Finally, it is worth noting that in the presented analysis, both observer and emitter are located at the equator. Therefore, although the full Kerr--MOG geodesic dynamics is Liouville integrable, the formalism does not rely on the polar part of the geodesic motion or on the explicit use of a Carter-type constant. This fact suggests that a similar strategy with suitable modifications can be applied to other stationary and axisymmetric non-Kerr geometries for which the full geodesic dynamics is not integrable. The key simplification is the restriction to equatorial circular motion, under which the formalism depends only on the conserved energy and axial angular momentum and does not require separability of the full geodesic equations.

\section{Discussion and final remarks}
\label{sec:conclusion}

In this paper, we derived exact analytic relations for the parameters of Schwarzschild--MOG and Kerr--MOG black holes in terms of directly measurable quantities associated with photons emitted by massive test particles on circular equatorial orbits. In the Schwarzschild--MOG case, the relevant observables are the total frequency shift, the telescope aperture angle, and the redshift rapidity. Evaluated on the midline and, when needed, close to the line of sight, these quantities allow one to obtain closed expressions for the MOG parameter \(\alpha\), the black hole mass \(M\), the distance to the observer \(D\), and the emitter radius \(r_e\). In this way, the static sector provides a fully analytic inversion of the observable relations without introducing auxiliary nonobservable parameters into the final formulas.

The contribution of modified gravity appears explicitly in the Schwarzschild--MOG observables through the parameter \(\alpha\). This makes it possible, at least in principle, to use the same observational framework not only for mass and distance determination, but also to test departures from the standard Schwarzschild geometry. In the limit \(\alpha\to 0\), all expressions reduce to the corresponding general relativistic results, as expected. We also showed that, in the weak-field regime, the redshift rapidity reduces to the line-of-sight projection of the Keplerian acceleration, with the MOG correction entering through the factor \(1+\alpha\).

We then extended the analysis to the rotating Kerr--MOG spacetime. In this case, the total frequency shift and the redshift rapidity are not sufficient to disentangle all the spacetime parameters, and one must also include the \emph{redshift acceleration}. Using these three observables on the midline, we obtained analytic relations that determine the spin parameter \(a\), together with \(M\), \(\alpha\), \(D\), and \(r_e\). Thus, the rotating case admits the same type of observable reconstruction as the static one, although with a richer structure due to frame dragging. In the corresponding weak-field limit, the \emph{redshift acceleration} reduces to the line-of-sight projection of the Keplerian jerk, while the dependence on the spin parameter appears only beyond leading order.

From the observational point of view, the formalism developed here is intended for real astrophysical systems that can be approximated by thin disks of emitters on nearly circular geodesic motion around supermassive black holes. In this sense, H$_2$O megamaser disks in AGNs remain a natural setting in which such relations are useful, since the emitting regions are often located at sub-parsec distances where a thin-disk treatment provides a reasonable approximation. At the same time, the present analysis does not include several effects that may be relevant in realistic environments close to the black hole, such as pressure gradients, magnetic fields, disk thickness, opacity, or deviations from exact circular motion. These effects become increasingly important as one approaches the central object and should be incorporated in future phenomenological applications.

To summarize, the main result of this work is the construction of closed, exact analytic formulas that express the parameters of Schwarzschild--MOG and Kerr--MOG black holes in terms of directly measurable redshift observables. In the static case, the method determines \(\alpha\), \(M\), \(D\), and \(r_e\); in the rotating case, it also determines the spin parameter \(a\) through the inclusion of the \emph{redshift acceleration}. Since the formulas reduce smoothly to the Schwarzschild and Kerr cases when \(\alpha=0\), they provide a natural extension of the redshift-based parameter-estimation program to black holes in Scalar--Tensor--Vector Gravity.

\section*{Acknowledgments}
MM acknowledges SNII and was supported by SECIHTI through Estancias Posdoctorales por M\'{e}xico Convocatoria
2023(1) under the postdoctoral Grant No. 1242413.

\appendix

\section{Redshift of non-geodesic emitters in MOG gravity}
\label{NONgeodesic}

In this Appendix, we present a brief discussion on the non-geodesic motion of massive test particles in a Schwarzschild-MOG black hole background, their frequency shift detected by a distant observer, and screened $\alpha$ in the weak-field limit due to the particle-vector field
interaction.

The equations of motion of massive test particles of mass $m$ and gravitational charge
$q=\kappa m=\sqrt{\alpha G_N} m$ are given by~\cite{Moffat2015EPJC}
\begin{equation}
\frac{d^2x^\mu}{d\tau^2}+\Gamma^\mu_{\alpha\beta}\frac{dx^\alpha}{d\tau}\frac{dx^\beta}{d\tau}=\kappa B_{\sigma}{}^\mu\frac{dx^\sigma}{d\tau},
\end{equation}
where satisfy the weak equivalence principle, and the right-hand side describes a fifth force arising from interactions between the massive particles and the vector field. In this scenario, the radial motion is described by~\cite{Turimov23}
\begin{eqnarray}
(U^r)^2&=&\left(E-\frac{\tilde q}{r}\right)^2-g(r)\left(1+\frac{L^2}{r^2}\right) \notag \\
&=&(E-V_+)(E-V_-),
\end{eqnarray}
where $\tilde q=\alpha M$ and $V_\pm$ is the effective potential
\begin{eqnarray}
V_\pm(r)=\frac{\tilde q}{r} \pm \sqrt{g(r)\left(1+\frac{L^2}{r^2}\right)}.
\label{Vplus}
\end{eqnarray}

On the other hand, the normalization condition $g_{\mu \nu}U^\mu U^\nu=-1$ for the equatorial circular orbits reduces to
\begin{equation}
(U^t_e)^2\left(-g_e+r_e^2\Omega^2\right)=-1,
\end{equation}
and therefore
\begin{equation}
U^t_e=\frac{1}{\sqrt{g_e-r_e^2\Omega^2}},\qquad U^\varphi_e=\frac{\Omega}{\sqrt{g_e-r_e^2\Omega^2}},
\label{UtvsOmega}
\end{equation}
where $\Omega=U^t_e/U^\varphi_e$ is the angular velocity of the emitter. Substituting this into the frequency-shift formula~\eqref{Redshift} gives the master expression
\begin{equation}
1+z=\frac{1-b_\varphi\Omega}{\sqrt{g_e-r_e^2\Omega^2}},
\label{NONgeoShift}
\end{equation}
in terms of the angular velocity. Since photons follow null geodesics due to their vanishing gravitational charge $q=0$, the fifth force vanishes and $b_\varphi$ is given by Eq.~\eqref{lbp}.

Now, for the geodesic motion of the photon source, $\Omega$ is given by the ratio of Eqs.~\eqref{Ut}-\eqref{Uphi}. But, for the non-geodesic motion, we compute $V_+^\prime=0$ using Eq.~\eqref{Vplus}, and then substitute $L=r_e^2\Omega U^t_e$ and $U^t_e$ from \eqref{UtvsOmega} to find
\begin{equation}
\Omega^2=\frac{1}{2r_e^2}\left[r_eg'_e-\frac{\tilde q^2}{r_e^2}-\frac{\tilde q}{r_e}\sqrt{\frac{\tilde q^2}{r_e^2}+4g_e-2r_eg'_e}\right],
\label{NONgeoOmega}
\end{equation}
where reduces to $\Omega^2=g'_e/(2r_e)$ for the geodesic motion in the limit $\tilde q \to 0$. 

Substituting $\Omega$ from ~\eqref{NONgeoOmega} into Eq.~\eqref{NONgeoShift} gives the final expression for the frequency shift of non-geodesic photon sources. In the Newtonian regime $M/r_e \to 0$, the redshift formula~\eqref{NONgeoShift} reduces to
\begin{equation}
z_{\text{Newton}} \approx \sqrt{\frac{M}{r_e}}\sin(\varphi+\delta)+\mathcal{O}\left( \frac{M}{r_e} \right),
\label{NONgeoWeakField}
\end{equation}
where shows that $\alpha$ is screened in the weak-field limit when considering the particle-vector field interaction [compare~\eqref{NONgeoWeakField} with Eq.~\eqref{RedshiftWeakField}].

\clearpage

\newpage

\begin{widetext}

\section{Decoupling the $5\times5$ system}
\label{appendix}

This appendix records a derivation of the closed midline solutions presented in Sec.~\ref{DecoupParam}. We restrict attention to the co-rotating branch and work on the midline $\varphi=\varphi_m=\pm\pi/2$, with $\delta=\delta_m$ and the redshift acceleration evaluated at $\pi/2 $ for simplicity (since the redshift rapidity only flips signs when evaluated at $\pm \pi/2 $,  it is taken as a single quantity, namely its absolute value). 

\subsection{The starting $5\times 5$ system}

At the midline, the system to be solved is given by~(\ref{eq:re})-(\ref{eq:zddotr})
\begin{subequations}\label{eq:appendix-start}
\begin{align}
r_e &= D\tan\delta_m, \label{eq:appendix-start-a}\\
1+z_{KMOG_1}^{(m)}
&=
\frac{1}{\sqrt{1-3\tilde P+2\tilde N+2\tilde a\sqrt{\tilde P-\tilde N}}}
\Biggl[
1+
\frac{\sqrt{\tilde P-\tilde N}}{1-2\tilde P+\tilde N}
\biggl(
\tilde a+
\frac{\tilde\Delta^{3/2}\cos\delta_m}
{\sqrt{\tilde\Delta^2\cos^2\delta_m+(1-2\tilde P+\tilde N)\sin^2\delta_m}}
\biggr)
\Biggr], \label{eq:appendix-start-b}\\
1+z_{KMOG_2}^{(m)}
&=
\frac{1}{\sqrt{1-3\tilde P+2\tilde N+2\tilde a\sqrt{\tilde P-\tilde N}}}
\Biggl[
1+
\frac{\sqrt{\tilde P-\tilde N}}{1-2\tilde P+\tilde N}
\biggl(
\tilde a-
\frac{\tilde\Delta^{3/2}\cos\delta_m}
{\sqrt{\tilde\Delta^2\cos^2\delta_m+(1-2\tilde P+\tilde N)\sin^2\delta_m}}
\biggr)
\Biggr], 
\label{eq:appendix-start-c}\\
\dot z_m
&=
\frac{D}{r_e}
\frac{(\tilde P-\tilde N)\tilde\Delta^{3/2}}
{\bigl(1+\tilde a\sqrt{\tilde P-\tilde N}\bigr)
\sqrt{1-3\tilde P+2\tilde N+2\tilde a\sqrt{\tilde P-\tilde N}}}
\left(\frac{D}{D^2+r_e^2}\right)
\frac{\sin\delta_m}
{\bigl(\tilde\Delta^2\cos^2\delta_m+(1-2\tilde P+\tilde N)\sin^2\delta_m\bigr)^{3/2}},
\label{eq:appendix-start-d}\\
\ddot{z}_{KMOG}^{(m)} &=
\left[
\frac{(\tilde{P} - \tilde{N})^{3/2}\tilde{\Delta}^{3/2}}
{
\left(1 + \tilde{a}\sqrt{\tilde{P} - \tilde{N}}\right)^{2}
\sqrt{
1 - 3\tilde{P} + 2\tilde{N}
+ 2\tilde{a}\sqrt{\tilde{P} - \tilde{N}}
}
}
\right] \notag
\\[0.75em]
&\times
\Bigg\{
\frac{D(D^{2} - r_{e}^{2})\sin\delta_{m}}
{
r_{e}(D^{2} + r_{e}^{2})^{2}
\left(
\tilde{\Delta}^{2}\cos^{2}\delta_{m}
+
(1 - 2\tilde{P} + \tilde{N})\sin^{2}\delta_{m}
\right)^{3/2}
}\notag
\\[0.75em]
&\qquad
-
\frac{
D^{4}\cos\delta_{m}
\left[
\tilde{\Delta}^{2}\cos^{2}\delta_{m}
+
\left(3\tilde{\Delta}^{2} - 2 + 4\tilde{P} - 2\tilde{N}\right)\sin^{2}\delta_{m}
\right]
}{
r_{e}^{2}(D^{2} + r_{e}^{2})^{2}
\left(
\tilde{\Delta}^{2}\cos^{2}\delta_{m}
+
(1 - 2\tilde{P} + \tilde{N})\sin^{2}\delta_{m}
\right)^{5/2}
}
\Bigg\},
\label{eq:appendix-start-e}
\end{align}
\end{subequations}
and we abbreviate
\begin{equation}\label{eq:smcm-defs}
s_m:=\sin\delta_m,
\qquad c_m:=\cos\delta_m, \qquad t_m=\frac{s_m}{c_m},
\end{equation}
and
\begin{equation}\label{eq:data-defs}
\begin{aligned}
\\
\dot z_m&:=\bigl|\dot z_{KMOG}^{(m)}\bigr|,
\end{aligned}
\end{equation}

\subsection{Compact midline form}

For the elimination, it is convenient to introduce the temporary combinations
\begin{equation}\label{eq:temp-aux}
X:=1-2\tilde P+\tilde N,
\qquad
Y:=\tilde P-\tilde N,
\qquad
Z:=1+\tilde a^{2}-2\tilde P+\tilde N=X+\tilde a^2,
\end{equation}
\begin{equation}\label{eq:temp-aux-2}
\mathcal E:=\sqrt{1-3\tilde P+2\tilde N+2\tilde a\sqrt Y},
\qquad
\Sigma:=\sqrt{Z^2c_m^2+Xs_m^2}.
\end{equation}
In terms of these blocks, the red/blue-shift equations yield immediately the half-sum and half-difference formulas
\begin{subequations}\label{eq:half-sum-diff}
\begin{align}
p_m &=  1+\frac{z_{KMOG_1}^{(m)}+z_{KMOG_2}^{(m)}}{2}  =     \frac{1}{\mathcal E}\left(1+\frac{\tilde a\sqrt Y}{X}\right), \label{eq:half-sum-diff-a}\\
d_m &= \frac{z_{KMOG_1}^{(m)}-z_{KMOG_2}^{(m)}}{2}  =      \frac{\sqrt Y\,Z^{3/2}c_m}{X\mathcal E\Sigma}. \label{eq:half-sum-diff-b}
\end{align}
\end{subequations}
Indeed, the term proportional to $\tilde\Delta^{3/2}\cos\delta_m=Z^{3/2} c_m$ cancels in the half-sum and doubles in the half-difference.

Next, using $r_e=D\tan\delta_m$, one has
\begin{equation}
\frac{D}{r_e}\frac{D}{D^2+r_e^2}\sin\delta_m
=
\frac{1}{\tan\delta_m}\frac{1}{D(1+\tan^2\delta_m)}\sin\delta_m
=
\frac{c_m^3}{D}.
\end{equation}
Hence Eq.~\eqref{eq:appendix-start-d} becomes
\begin{equation}\label{eq:compact-rapidity-a}
\dot z_m
=
\frac{YZ^{3/2}c_m^3}{D\bigl(1+\tilde a\sqrt Y\bigr)\mathcal E\Sigma^3}.
\end{equation}
Now for simplification purposes, we also introduce the new quantity $\eta_m =  -\ddot z_{KMOG}^{(m)}/\dot z_m^{\,2}$, 

\begin{equation}\label{eq:compact-rapidity-b}
\eta_m
=
\frac{\mathcal E\Sigma}{\sqrt Y\,Z^{3/2}s_m^2c_m^3}
\Bigl[Z^2c_m^4+(Z^2-2X)s_m^2c_m^2+Xs_m^4\Bigr].
\end{equation}
Equations \eqref{eq:half-sum-diff}, \eqref{eq:compact-rapidity-a}, and \eqref{eq:compact-rapidity-b} are algebraically equivalent to the original relations \eqref{eq:appendix-start-b}--\eqref{eq:appendix-start-e}, but they are much better suited for elimination.

\subsection{First elimination}

The first useful observation is that the product of \eqref{eq:half-sum-diff-b} and \eqref{eq:compact-rapidity-b} is free of $\mathcal E$, $\Sigma$, $Y$, and $Z^{3/2}$. Multiplying the two relations gives
\begin{align}
d_m\eta_m
&=
\left(\frac{\sqrt Y\,Z^{3/2}c_m}{X\mathcal E\Sigma}\right)
\left(
\frac{\mathcal E\Sigma}{\sqrt Y\,Z^{3/2}s_m^2c_m^3}
\Bigl[Z^2c_m^4+(Z^2-2X)s_m^2c_m^2+Xs_m^4\Bigr]
\right) \notag\\
&=
\frac{Z^2c_m^4+(Z^2-2X)s_m^2c_m^2+Xs_m^4}{Xs_m^2c_m^2}.
\end{align}
Splitting the fraction term by term,
\begin{align}
d_m\eta_m
&=
\frac{Z^2c_m^4}{Xs_m^2c_m^2}
+
\frac{(Z^2-2X)s_m^2c_m^2}{Xs_m^2c_m^2}
+
\frac{Xs_m^4}{Xs_m^2c_m^2} \notag\\
&=
\frac{Z^2c_m^2}{Xs_m^2}
+
\frac{Z^2}{X}-2+\frac{s_m^2}{c_m^2} \notag\\
&=
\frac{Z^2}{X}\left(\frac{c_m^2}{s_m^2}+1\right)-2+\tan^2\delta_m \notag\\
&=
\frac{Z^2}{Xs_m^2}-2+\tan^2\delta_m, \label{eq:dm-eta-relation}
\end{align}
and rearranging leads
\begin{equation}\label{eq:z2overx-first}
\frac{Z^2}{X}=s_m^2\bigl(d_m\eta_m+2-\tan^2\delta_m\bigr).
\end{equation}
At this stage, in order to simplify the expression we introduce the quantities
\begin{equation}\label{eq:HK-defs}
H_m:=3+d_m\eta_m,
\qquad
K_m:=\tan^2\delta_m\bigl(c_m^2H_m-1\bigr).
\end{equation}
Now, $K_m$ can be rewritten as follows
\begin{align}
K_m
&=
\frac{s_m^2}{c_m^2}\bigl(c_m^2(3+d_m\eta_m)-1\bigr) \notag\\
&=
3s_m^2+s_m^2d_m\eta_m-\tan^2\delta_m \notag\\
&=
s_m^2\bigl(d_m\eta_m+2-\tan^2\delta_m\bigr),
\end{align}
and comparison with \eqref{eq:z2overx-first} yields the key identity
\begin{equation}\label{eq:key-k-ratio}
K_m=\frac{Z^2}{X},
\end{equation}
or equivalently,
\begin{equation}\label{eq:x-z-k}
Z^2=XK_m.
\end{equation}

\subsection{Solving for the reduced system}

To solve the system explicitly, we introduce one temporary parameter,
\begin{equation}\label{eq:phi-def}
\Phi_m:=\frac{Z}{K_m}.
\end{equation}
Then \eqref{eq:key-k-ratio} implies
\begin{equation}\label{eq:xzphi}
X=K_m\Phi_m^2,
\qquad Z=K_m\Phi_m.
\end{equation}
Inserting these into the definition of $\Sigma$ gives
\begin{align}
\Sigma^2
&=Z^2c_m^2+Xs_m^2 \notag\\
&=(K_m\Phi_m)^2c_m^2+(K_m\Phi_m^2)s_m^2 \notag\\
&=K_m\Phi_m^2\bigl(K_mc_m^2+s_m^2\bigr).
\end{align}
Using
\begin{equation}
K_m=\tan^2\delta_m(c_m^2H_m-1)
\quad\Longrightarrow\quad
K_mc_m^2+s_m^2=s_m^2 c_m^2H_m,
\end{equation}
we obtain
\begin{equation}\label{eq:sigma-simplified}
\Sigma^2=K_m\Phi_m^2s_m^2c_m^2H_m,
\qquad
\Sigma=\sqrt{K_m}\,\Phi_m s_mc_m\sqrt{H_m}.
\end{equation}
Substituting \eqref{eq:xzphi} and \eqref{eq:sigma-simplified} into \eqref{eq:half-sum-diff-b} gives
\begin{align}
d_m
&=
\frac{\sqrt Y\,(K_m\Phi_m)^{3/2}c_m}
{(K_m\Phi_m^2)\mathcal E\,(\sqrt{K_m}\,\Phi_m s_mc_m\sqrt{H_m})} \notag\\
&=
\frac{\sqrt Y}{\mathcal E\,s_m\sqrt{H_m}\,\Phi_m^{3/2}}.
\end{align}
Therefore,
\begin{equation}\label{eq:sqrtY-formula}
\sqrt Y=d_m\mathcal E s_m\sqrt{H_m}\,\Phi_m^{3/2}.
\end{equation}
On the other hand, since $Z=X+\tilde a^2$, equations \eqref{eq:xzphi} imply
\begin{equation}\label{eq:atilde-squared}
\tilde a^2=Z-X=K_m\Phi_m(1-\Phi_m).
\end{equation}
For the co-rotating branch we take the positive root,
\begin{equation}\label{eq:atilde-positive}
\tilde a=\sqrt{K_m\Phi_m(1-\Phi_m)}.
\end{equation}

\subsubsection{Determining $\mathcal E$ and $\Phi_m$}

Equation \eqref{eq:half-sum-diff-a} can be rewritten as
\begin{equation}\label{eq:aq-from-p}
\tilde a\sqrt Y=X(p_m\mathcal E-1).
\end{equation}
Substituting \eqref{eq:xzphi}, \eqref{eq:sqrtY-formula}, and \eqref{eq:atilde-positive} into \eqref{eq:aq-from-p}, we obtain
\begin{align}
K_m\Phi_m^2(p_m\mathcal E-1)
&=
\sqrt{K_m\Phi_m(1-\Phi_m)}
\cdot d_m\mathcal E s_m\sqrt{H_m}\,\Phi_m^{3/2} \notag\\
&=d_m\mathcal E s_m\sqrt{H_m}\,\Phi_m^2\sqrt{K_m(1-\Phi_m)}.
\end{align}
After dividing by $K_m\Phi_m^2$ one finds
\begin{equation}\label{eq:pE-minus-one}
p_m\mathcal E-1=d_m\mathcal E\,R_m,
\qquad
R_m:=s_m\sqrt{\frac{H_m(1-\Phi_m)}{K_m}}.
\end{equation}
Hence
\begin{equation}\label{eq:E-formula-temp}
\mathcal E=\frac{1}{p_m-d_mR_m}.
\end{equation}
The remaining task is to determine $\Phi_m$. Starting from the definition of $\mathcal E$,
\begin{equation}
\mathcal E^2=X-Y+2\tilde a\sqrt Y,
\end{equation}
and using \eqref{eq:aq-from-p}, we get
\begin{equation}
\mathcal E^2=X-Y+2X(p_m\mathcal E-1).
\end{equation}
Substituting \eqref{eq:xzphi}, \eqref{eq:sqrtY-formula}, and \eqref{eq:pE-minus-one}, this becomes
\begin{equation}
\mathcal E^2
=K_m\Phi_m^2-d_m^2\mathcal E^2s_m^2H_m\Phi_m^3+2K_m\Phi_m^2d_m\mathcal ER_m.
\end{equation}
Replacing $\mathcal E$ by \eqref{eq:E-formula-temp} and multiplying by $(p_m-d_mR_m)^2$ gives
\begin{align}
1
&=K_m\Phi_m^2(p_m-d_mR_m)^2-d_m^2s_m^2H_m\Phi_m^3
+2K_m\Phi_m^2d_mR_m(p_m-d_mR_m) \notag\\
&=K_m\Phi_m^2(p_m^2-d_m^2R_m^{2})-d_m^2s_m^2H_m\Phi_m^3.
\end{align}
Using the definition of $R_m$,
\begin{equation}\label{eq:A34}
R_m^2=s_m^2\frac{H_m(1-\Phi_m)}{K_m},
\end{equation}
we obtain
\begin{align}
1
&=K_m\Phi_m^2p_m^2-d_m^2s_m^2H_m\Phi_m^2(1-\Phi_m)-d_m^2s_m^2H_m\Phi_m^3 \notag\\
&=\Phi_m^2\bigl(K_mp_m^2-d_m^2s_m^2H_m\bigr).
\end{align}
Therefore,
\begin{equation}\label{eq:Phi-solved}
\Phi_m=\frac{1}{\sqrt{K_mp_m^2-d_m^2s_m^2H_m}}.
\end{equation}
From Equation (\ref{eq:A34})
\begin{equation}\label{eq:F-identification}
R_m=s_m\sqrt{\frac{H_m(1-\Phi_m)}{K_m}}.
\end{equation}

Likewise, \eqref{eq:E-formula-temp} becomes
\begin{equation}\label{eq:E-solved}
E_m:=\mathcal E=\frac{1}{p_m-d_mR_m}.
\end{equation}

\subsubsection{Reconstruction of the reduced variables}

With $\Phi_m$ and $R_m$ in hand, the remaining reduced quantities follow directly:
\begin{subequations}\label{eq:reduced-vars}
\begin{align}
\mathcal A_m:=1-2\tilde P+\tilde N &= X = K_m\Phi_m^2 = \frac{K_m}{K_mp_m^2-d_m^2s_m^2H_m}, \label{eq:reduced-vars-a}\\
\tilde\Delta_m:=1+\tilde a^2-2\tilde P+\tilde N &= Z = K_m\Phi_m = \frac{K_m}{\sqrt{K_mp_m^2-d_m^2s_m^2H_m}}, \label{eq:reduced-vars-b}\\
\tilde a_m &= \sqrt{K_m\Phi_m(1-\Phi_m)}, \label{eq:reduced-vars-c}\\
q_m^2:=\tilde P-\tilde N &= Y = d_m^2s_m^2H_m\Phi_m^3E_m^2
=\frac{d_m^2s_m^2H_m\Phi_m^3}{(p_m-d_mR_m)^2}. \label{eq:reduced-vars-d}
\end{align}
\end{subequations}
Since
\begin{equation}
q_m^2=\tilde P-\tilde N,
\qquad
\mathcal A_m=1-2\tilde P+\tilde N,
\end{equation}
we solve for the dimensionless effective potentials as
\begin{subequations}\label{eq:PN-solved}
\begin{align}
\tilde P_m &= 1-\mathcal A_m-q_m^2, \label{eq:PN-solved-a}\\
\tilde N_m &= 1-\mathcal A_m-2q_m^2. \label{eq:PN-solved-b}
\end{align}
\end{subequations}

\subsection{Distance, source radius, spin, MOG coupling, and mass}

Starting from \eqref{eq:compact-rapidity-a}, we substitute $Y=q_m^2$, $Z=K_m\Phi_m$, $\Sigma=\sqrt{K_m}\Phi_ms_mc_m\sqrt{H_m}$, and $E_m=1/(p_m-d_mR_m)$, and after a short simplification, it gives
\begin{equation}\label{eq:dotz-simplified}
\dot z_m
=
\frac{d_m^2E_m\Phi_m^{3/2}}{D(1+\tilde a_m q_m)s_m\sqrt{H_m}}.
\end{equation}
Now, we note that
\begin{equation}
\tilde a_m q_m=\mathcal A_m(p_mE_m-1)=\mathcal A_m d_mE_mR_m,
\end{equation}
where the second equality follows from \eqref{eq:pE-minus-one}. Hence
\begin{equation}
\frac{E_m}{1+\tilde a_m q_m}
=
\frac{1}{p_m+d_mR_m(\mathcal A_m-1)}.
\end{equation}
Substituting this into \eqref{eq:dotz-simplified} yields the closed expression for the distance
\begin{equation}\label{eq:D-paper-form}
D=
\frac{d_m^2\Phi_m^{3/2}}
{\dot z_m s_m\sqrt{H_m}\,[p_m+d_mR_m(\mathcal A_m-1)]}.
\end{equation}
The remaining observables are then obtained from
\begin{subequations}\label{eq:physical-observables}
\begin{align}
r_e &= D\tan\delta_m, \label{eq:physical-observables-a}\\
a &= \tilde a_m r_e, \label{eq:physical-observables-b}\\
\alpha &= \frac{\tilde N_m}{\tilde P_m^2-\tilde N_m}, \label{eq:physical-observables-c}\\
M &= r_e\left(\tilde P_m-\frac{\tilde N_m}{\tilde P_m}\right). \label{eq:physical-observables-d}
\end{align}
\end{subequations}
Equations \eqref{eq:HK-defs},  \eqref{eq:Phi-solved}, \eqref{eq:F-identification}, \eqref{eq:reduced-vars}, \eqref{eq:PN-solved}, \eqref{eq:D-paper-form}, and \eqref{eq:physical-observables} are the decoupled solution of the system in terms of the reduced variables.

\subsection{Explicit formulas}

Now, by substituting the reduced variables appearing in the decoupled solutions directly in terms of the initially defined parameters  $p_m, d_m, s_m, c_m, \dot{z}_m, \eta_m$, we found that it was also convenient to introduce the additional parameter due to its recurrent appearance in the expanded solutions 
\begin{equation}\label{eq:Lambda-def}
\Lambda_m:=K_mp_m^2-d_m^2s_m^2H_m.
\end{equation}
By considering~\eqref{eq:HK-defs}, $\Lambda_m$ can be written more explicitly as follows
\begin{equation}\label{eq:Lambda-expanded}
\Lambda_m
=
\frac{s_m^2}{c_m^2}\Bigl(c_m^2(3+d_m\eta_m)-1\Bigr)p_m^2
-d_m^2s_m^2(3+d_m\eta_m).
\end{equation}
One also notes that the explicit forms of $\Phi_m$, $R_m$, $\mathcal A_m$, and $q_m^2$ are given by
\begin{equation}\label{eq:F-R-explicit}
\Phi_m=\Lambda_m^{-1/2},
\qquad
R_m=s_m\sqrt{\frac{(3+d_m\eta_m)\bigl(1-\Lambda_m^{-1/2}\bigr)}{t_{m} ^2\Bigl(c_m^2(3+d_m\eta_m)-1\Bigr)}}.
\end{equation}
\begin{equation}\label{eq:A-explicit}
\mathcal A_m
=
\frac{t_{m} ^2\Bigl(c_m^2(3+d_m\eta_m)-1\Bigr)}{\Lambda_m},
\qquad
q_m^2
=
\frac{d_m^2s_m^2(3+d_m\eta_m)\Lambda_m^{-3/2}}
{\left[p_m-d_ms_m\sqrt{\dfrac{(3+d_m\eta_m)(1-\Lambda_m^{-1/2})}{t_{m} ^2\bigl(c_m^2(3+d_m\eta_m)-1\bigr)}}\right]^2}.
\end{equation}
Substituting \eqref{eq:F-R-explicit}--\eqref{eq:A-explicit} into the compact solutions~\eqref{eq:D-paper-form}--\eqref{eq:physical-observables} and simplifying gives the explicit formulas

{\small
\begin{equation}
\begin{aligned}
D
&=
\frac{d_m^2\Lambda_m^{-3/4}}
{\dot z_m\,s_m\,\sqrt{3+d_m\eta_m}}
\left[
p_m
+
d_m s_m
\sqrt{
\dfrac{(3+d_m\eta_m)\bigl(1-\Lambda_m^{-1/2}\bigr)}
{t_{m} ^2\bigl(c_m^2(3+d_m\eta_m)-1\bigr)}
}
\Biggl(
\dfrac{s_m^2\bigl(c_m^2(3+d_m\eta_m)-1\bigr)}{c_m^2\Lambda_m}
-1
\Biggr)
\right]^{-1},
\label{Dapp}
\end{aligned}
\end{equation}

\begin{equation}
\begin{aligned}
r_e
&=
\frac{d_m^2\Lambda_m^{-3/4}}
{\dot z_m\,c_m\,\sqrt{3+d_m\eta_m}}
\left[
p_m
+
d_m s_m
\sqrt{
\dfrac{(3+d_m\eta_m)\bigl(1-\Lambda_m^{-1/2}\bigr)}
{t_{m} ^2\bigl(c_m^2(3+d_m\eta_m)-1\bigr)}
}
\Biggl(
\dfrac{s_m^2\bigl(c_m^2(3+d_m\eta_m)-1\bigr)}{c_m^2\Lambda_m}
-1
\Biggr)
\right]^{-1},
\end{aligned}
\end{equation}

\begin{equation}
\begin{aligned}
\alpha
&=
\left[
1
-
\dfrac{s_m^2\Bigl(c_m^2(3+d_m\eta_m)-1\Bigr)}{c_m^2\Lambda_m}
-
\dfrac{2d_m^2s_m^2(3+d_m\eta_m)\Lambda_m^{-3/2}}
{\Biggl[
p_m-d_ms_m\sqrt{\dfrac{(3+d_m\eta_m)(1-\Lambda_m^{-1/2})}{t_{m} ^2\bigl(c_m^2(3+d_m\eta_m)-1\bigr)}}\Biggr]^2}
\right]
\\
&\quad\times
\left\{
\left[
1
-
\dfrac{s_m^2\Bigl(c_m^2(3+d_m\eta_m)-1\Bigr)}{c_m^2\Lambda_m}
-
\dfrac{d_m^2s_m^2(3+d_m\eta_m)\Lambda_m^{-3/2}}
{\Biggl[
p_m-d_ms_m\sqrt{\dfrac{(3+d_m\eta_m)(1-\Lambda_m^{-1/2})}{t_{m} ^2\bigl(c_m^2(3+d_m\eta_m)-1\bigr)}}\Biggr]^2}
\right]^2 \right.
\\
&\qquad-
\left. \left[
1
-
\dfrac{s_m^2\Bigl(c_m^2(3+d_m\eta_m)-1\Bigr)}{c_m^2\Lambda_m}
-
\dfrac{2d_m^2s_m^2(3+d_m\eta_m)\Lambda_m^{-3/2}}
{\Biggl[
p_m-d_ms_m\sqrt{\dfrac{(3+d_m\eta_m)(1-\Lambda_m^{-1/2})}{t_{m} ^2\bigl(c_m^2(3+d_m\eta_m)-1\bigr)}}\Biggr]^2}
\right]
\right\}^{-1},
\end{aligned}
\end{equation}

\begin{equation}
\begin{aligned}
M
&=
\frac{d_m^2\Lambda_m^{-3/4}}
{\dot z_m\,c_m\,\sqrt{3+d_m\eta_m}}
\left[
p_m
+
d_m s_m
\sqrt{
\dfrac{(3+d_m\eta_m)\bigl(1-\Lambda_m^{-1/2}\bigr)}
{t_{m} ^2\bigl(c_m^2(3+d_m\eta_m)-1\bigr)}
}
\Biggl(
\dfrac{s_m^2\bigl(c_m^2(3+d_m\eta_m)-1\bigr)}{c_m^2\Lambda_m}
-1
\Biggr)
\right]^{-1}
\\
&\quad\times
\left[
\left(
1
-
\dfrac{s_m^2\Bigl(c_m^2(3+d_m\eta_m)-1\Bigr)}{c_m^2\Lambda_m}
-
\dfrac{d_m^2s_m^2(3+d_m\eta_m)\Lambda_m^{-3/2}}
{\Biggl[
p_m-d_ms_m\sqrt{\dfrac{(3+d_m\eta_m)(1-\Lambda_m^{-1/2})}{t_{m} ^2\bigl(c_m^2(3+d_m\eta_m)-1\bigr)}}\Biggr]^2}
\right) \right.
\\
&\qquad \left.-
\dfrac{
\left(
1
-
\dfrac{s_m^2\Bigl(c_m^2(3+d_m\eta_m)-1\Bigr)}{c_m^2\Lambda_m}
-
\dfrac{2d_m^2s_m^2(3+d_m\eta_m)\Lambda_m^{-3/2}}
{\Biggl[
p_m-d_ms_m\sqrt{\dfrac{(3+d_m\eta_m)(1-\Lambda_m^{-1/2})}{t_{m} ^2\bigl(c_m^2(3+d_m\eta_m)-1\bigr)}}\Biggr]^2}
\right)
}{
\left(
1
-
\dfrac{s_m^2\Bigl(c_m^2(3+d_m\eta_m)-1\Bigr)}{c_m^2\Lambda_m}
-
\dfrac{d_m^2s_m^2(3+d_m\eta_m)\Lambda_m^{-3/2}}
{\Biggl[
p_m-d_ms_m\sqrt{\dfrac{(3+d_m\eta_m)(1-\Lambda_m^{-1/2})}{t_{m} ^2\bigl(c_m^2(3+d_m\eta_m)-1\bigr)}}\Biggr]^2}
\right)
}
\right],
\end{aligned}
\end{equation}

\begin{equation}
\begin{aligned}
a
&=
\frac{d_m^2\Lambda_m^{-3/4}}
{\dot z_m\,c_m\,\sqrt{3+d_m\eta_m}} \sqrt{
t_{m} ^2\bigl(c_m^2(3+d_m\eta_m)-1\bigr)
\Lambda_m^{-1/2}
\bigl(1-\Lambda_m^{-1/2}\bigr)
}
\\
&\quad\times
\left[
p_m
+
d_m s_m
\sqrt{
\dfrac{(3+d_m\eta_m)\bigl(1-\Lambda_m^{-1/2}\bigr)}
{t_{m} ^2\bigl(c_m^2(3+d_m\eta_m)-1\bigr)}
}
\Biggl(
\dfrac{s_m^2\bigl(c_m^2(3+d_m\eta_m)-1\bigr)}{c_m^2\Lambda_m}
-1
\Biggr)
\right]^{-1},
\label{aapp}
\end{aligned}
\end{equation}
}
where we recall the following shorthand notations incorporated in the aforementioned formulas for convenience
\begin{equation}
s_m:=\sin\delta_m,
\qquad
c_m:=\cos\delta_m,
\qquad
t_m:= \tan \delta_m,
\end{equation}

\begin{equation}
p_m:=1+\frac{z_{KMOG_{1}}^{(m)}+z_{KMOG_{2}}^{(m)}}{2},
\qquad
d_m:=\frac{z_{KMOG_{1}}^{(m)}-z_{KMOG_{2}}^{(m)}}{2},
\qquad
\dot z_m:=\bigl|\dot z_{KMOG}^{(m)}\bigr|,
\qquad
\eta_m:=-\frac{\ddot z_{KMOG}^{(m)}}{\dot z_m^{\,2}} ,
\end{equation}

\begin{equation}
\Lambda_m
=
\frac{s_m^2}{c_m^2}\Bigl(c_m^2(3+d_m\eta_m)-1\Bigr)p_m^2
-d_m^2s_m^2(3+d_m\eta_m).
\end{equation}

Equations (\ref{Dapp})--(\ref{aapp}) provide the expanded solutions of the system, and they represent exact analytic formulas for the spacetime variables $\{M,a,\alpha,D,r_e\}$ in terms of purely observational quantities $\left\{z_{KMOG_{1}}^{(m)},z_{KMOG_{2}}^{(m)},\dot z_{KMOG}^{(m)},\ddot z_{KMOG}^{(m)},\delta_m\right\}$. The counter-rotating branch is obtained by repeating the same elimination procedure with the opposite sign choice for  $\tilde{a}$.

\end{widetext}

\end{document}